\documentstyle[12pt]{article}
\input psfig.sty
\newlength{\captsize} \let\captsize=\small 
\begin{document}
\newcommand\ie {{\it i.e. }}
\newcommand\eg {{\it e.g. }}
\newcommand\etc{{\it etc. }}
\newcommand\cf {{\it cf.  }}
\newcommand\etal {{\it et al. }}
\def\mw{m_W}
\def\mz{m_Z}
\def\to{\rightarrow}
\def\anti{\overline}
\def\lsim{\mathrel{\raise.3ex\hbox{$<$\kern-.75em\lower1ex\hbox{$\sim$}}}}
\def\gsim{\mathrel{\raise.3ex\hbox{$>$\kern-.75em\lower1ex\hbox{$\sim$}}}}
\def\epem{e^+e^-}
\def\mupmum{\mu^+\mu^-}
\def\anti{\overline}
\def\pbi{~{\rm pb}^{-1}}
\def\fbi{~{\rm fb}^{-1}}
\def\nbi{~{\rm nb}^{-1}}
\def\fb{~{\rm fb}}
\def\pb{~{\rm pb}}
\def\ev{\,{\rm eV}}
\def\mev{\,{\rm MeV}}
\def\gev{\,{\rm GeV}}
\def\tev{\,{\rm TeV}}
\def\rts{{\sqrt s}}
\def\gam{\gamma}
\def\lamqcd{\Lambda_{\rm QCD}}

\textwidth6in
\textheight8.5in
\oddsidemargin.25in
\topmargin.25in
\headheight0in
\headsep0in

\begin{flushright}
UCD-96-31    \\
LBNL-39474  \\
October, 1996 
\end{flushright} 
\vskip 0.2cm
\begin{center}
{\Large \bf Determining the existence and nature of
the quark-gluon plasma by Upsilon suppression at the LHC} 
\end{center}
\vskip 0.2cm
\centerline{\sc J.F. Gunion$^{\dagger}$\footnotetext{$^\dagger$Supported
by the Department of Energy under Contract No. DE-FG03-91ER40674
 and by the Davis Institute for High Energy Physics.}}
\vskip 0.2cm
\centerline{\it Davis Institute for High Energy Physics, 
Physics Department,}
\centerline{\it University of California, Davis, California 95616 USA}
\vskip 0.2cm
\centerline{\sc R. Vogt$^{\star}$\footnotetext{$^\star$This work was supported
in part by the Director, Office of Energy Research, Division of Nuclear Physics
of the Office of High Energy and Nuclear Physics of the U. S.
Department of Energy under Contract No. DE-AC03-76SF0098.}}
\vskip 0.2cm
\centerline{\it Nuclear Science Division,}
\centerline{\it Lawrence Berkeley National Laboratory, 
Berkeley, California 94720}
\centerline{\it and}
\centerline{\it Physics Department,}
\centerline{\it University of California, Davis, California 95616 USA}

\vskip 1.cm
\centerline{\bf Abstract}
\vskip 0.6cm
{\small
Minijet production in $\sqrt{s} = 5.5$ TeV Pb+Pb collisions at the LHC is
expected to produce a gluon-dominated plasma with large initial temperatures
and energy densities.
We discuss the implications of the high initial temperatures on the screening
mass and on the relative suppression of members of the $\Upsilon$ family.
We demonstrate that $p_T$-dependence of the $\Upsilon'/\Upsilon$
ratio is only expected if a quark-gluon plasma with significant
energy density is present, and that the precise dependence of this
ratio will strongly constrain the QGP models.
We also propose that the $p_T$ spectra of the $\Upsilon$ resonances
themselves will provide valuable information by comparing these spectra
to that of the $Z$ boson, which should not be influenced by plasma production.}
\vfill

\newpage

\section{Introduction}

\indent\indent 
Finite temperature simulations of lattice gauge theory suggest a phase
transition at high energy density and temperature
to a new phase of QCD matter --- the quark-gluon plasma (QGP).
In many models, the temperatures and energy densities reached
in central nucleus-nucleus collisions at 
the Brookhaven Relativistic Heavy Ion Collider (RHIC) and the 
Large Hadron Collider (LHC) at CERN are predicted
to be sufficient for QGP formation.
Demonstrating that such a QGP has formed and understanding its
nature will be a very important component of the experimental programs
at both machines. In this paper, we focus on using measurements
of the $p_T$ spectra of the Upsilon resonance family 
to accomplish these tasks, emphasizing the possible advantages
of the higher LHC energy as compared to RHIC energy.

The details of the QCD phase transition to the QGP, as obtained in finite
temperature simulations of lattice gauge theory,     
are strongly dependent on the number of quark
flavors, $n_f$, and the number of colors, $N$.
The critical temperature, $T_c$, and the order
of the phase transition are particularly sensitive to the values of
$n_f$ and $N$ used in the simulations \cite{Karsch}.
In a pure SU($N$) gauge theory, $n_f=0$, the phase 
transition is second order, {\it i.e.} continuous, for
SU(2) and first order for SU(3).  For the SU($N$) theory, the critical
temperature, $T_c$, is 260 MeV.  When dynamical light quarks are
included in the lattice simulation, 
the critical temperature is substantially lower and the order
of the phase transition can be altered: for $n_f=2$,
$T_c \approx 150$ MeV and the phase transition
appears to be continuous; for simulations with more light flavors, 
$n_f \geq 3$, $T_c$ is unaltered but
the transition again becomes first order.  This latter
conclusion depends on the relative quark masses \cite{Brown}.
The critical energy density at $T_c$ is only weakly dependent on the color
and flavor degrees of freedom; $\epsilon_c\approx 1-2\gev$/fm$^3$ is obtained 
both with and without quark degrees of freedom for $N=2$ or 3. 
For a QGP to be formed in an ultrarelativistic heavy-ion collision, 
the initial temperature, $T_0$, and energy 
density, $\epsilon(t_0)$, of the quark-gluon system
created by the collision must be larger than $T_c$ and $\epsilon_c$.

Part of the LHC experimental program will be devoted to heavy ion collisions
such as Pb+Pb at $\sqrt{s} = 5.5$ TeV.  At this high energy, the highest
available for heavy-ion collisions, perturbative QCD processes will be
the dominant factor in determining the initial state of the post-collision
quark/gluon system. In particular, at early times,
$t_0 \sim 1/p_T\leq 1/p_0$,  semihard 
production of minijets\footnote{Minijets are jets with $p_T\ge p_0\sim 1-3$ 
GeV \cite{GH}, usually not observable as individual jets below $p_T 
\sim 5$ GeV \cite{UA1}.} could be dominant in establishing the 
initial conditions for further evolution of the system \cite{EG}.  
The appropriate value of $p_0$ is uncertain but should be of order 
$p_0\sim 2\gev$, in which case $t_0 \sim 0.1$ fm. The crucial
quantity in determining whether a quark-gluon plasma forms is the initial
energy density, given by
\begin{equation}
\epsilon(t_0)= { \overline E_T p_0\over \pi R^2}\,,~{\rm with}~\overline 
E_T=T_{\rm AA}({\bf 0})\sigma_{\rm jet} \langle E_T\rangle
\,.
\label{epsform}
\end{equation}
Here, $R$ is of order the nuclear radius, $R_A$,
$T_{\rm AA}({\bf 0})$ is the nuclear overlap 
for a central collision\footnote{The
nuclear overlap function for a collision of projectile $A$ with target $B$ is
$T_{AB}({\bf b}) = \int d^2 {\bf s} dz dz' \rho_A({\bf s},z) \rho_B({\bf
b}-{\bf s},z')$ where 
the impact parameter ${\bf b}$ is zero in the most central collisions.} and
$\sigma_{\rm jet}\langle E_T\rangle$ is the first $p_T$ moment of the minijet
production cross section in the central unity of rapidity in $pp$ collisions
 (if we neglect nuclear shadowing) 
using the assumed value of $p_0$. The initial temperature $T_0$
is proportional to $[\epsilon(t_0)]^{1/4}$. For given $p_0$,
$\sigma_{\rm jet}\langle E_T\rangle$ is computable from QCD once
the parton distribution functions (PDF's) for the colliding protons
and the types of jets participating in the plasma are both specified.
If we take $p_0=2\gev$, employ MRS D$-^\prime$ \cite{MRS}
parton densities and include only gluon jets (\ie assume
a pure gluon plasma) with rapidity
$|y|\leq 0.5$, then the semihard production of minijets
leads to a large initial energy density, $\epsilon(t_0)=1170$
GeV/fm$^3$, and initial temperature, $T_0 = 1.14$ GeV, 
in central Pb+Pb collisions at $\rts=5.5\tev$, to be compared
to $\epsilon(t_0)=27$ GeV/fm$^3$ and $T_0=0.46$ GeV at the lower RHIC energy.
See \cite{EKR,KJE95} for details.  

The sensitivity of the critical parameters
to the choice of PDF set is illustrated
in Table~\ref{sensitivity}, following \cite{EKR,KJE95}. 
MRS D$-^\prime$ provides the most
optimistic initial temperature, with the MRS H distributions \cite{mrsh} 
(designed to fit early HERA data) not far behind.  The
Duke-Owens-I (DO1) \cite{dukeowens},
and MRS D0$^\prime$ \cite{MRS} distributions that grow more slowly
as $x$ decreases yield smaller initial energy density. More recent HERA data
continues to favor distributions (such as the MRS A parameterization 
\cite{mrsa}) that at small $x$ are similar to the H
parameterization, \ie that do not grow quite so rapidly at small-$x$ 
as the D$-^\prime$ distributions but are certainly much more
singular than either DO1 or D0$^\prime$. 
All the numbers quoted in the table will become smaller when
reductions in parton density due to shadowing in the nucleus are incorporated.

\begin{table}[htb]
\begin{center}
\begin{tabular}{|c|c|c|c|c|c|} \hline
PDF & \#jets & \# gluon jets & $\overline E_T$ (GeV) &  $T_0$ (GeV) & 
$dN_{\rm ch}^{\rm gluons~ only}/dy$ \\ \hline
DO1 & 776 & 640 & 3200 & 0.74 & 914 \\ \hline
D0$^\prime$ & 1510 & 1250 & 5440 & 0.81 & 1350 \\ \hline
H & 3250 & 2710 & 10300 & 0.96 & 2130 \\ \hline
D$-^\prime$ & 5980 & 5220 & 17600 & 1.14 & 3360 \\ \hline
\end{tabular}
\end{center}
\caption[]{Dependence of initial temperature and particle density
on distribution function choice at $\rts=5.5\tev$. The results are taken
from Ref.\ \cite{KJE95}.  We employ
$T_{\rm PbPb}({\bf 0})=32~{\rm mb}^{-1}$. The distributions employed
are Duke-Owens I, DO1 \cite{dukeowens}, the 
MRS  D0$^\prime$ and D$-^\prime$ from \cite{MRS}, and the MRS H distributions
from \cite{mrsh}.}
\label{sensitivity}
\end{table}

The sensitivity of the results to $p_0$ depends upon the PDFs.
Increasing $p_0$ from 2 to 4 GeV
decreases $\sigma_{\rm jet} \langle E_T\rangle$ by factors
of 5, 3, 2, and 1.5 for D$-^\prime$ \cite{MRS}, 
H \cite{mrsh}, D0$^\prime$ \cite{MRS}, and DO1 
\cite{dukeowens} distributions,
respectively, at $\rts=5.5\tev$. However,
$\epsilon(t_0)$, eq.~(\ref{epsform}), also proportional to $p_0$,
changes by a factor of two or less, even increasing in the DO1 case.
The sensitivity of $\epsilon(t_0)$ to $p_0$ at RHIC is much greater, declining
by as much as a factor of 5 between $p_0=2\gev$ and $p_0=4\gev$.
Inclusion of quark as well as gluon jets would increase $\epsilon(t_0)$
by roughly 30\% at $\rts=5.5\tev$, but if the plasma is a quark-gluon
plasma with 2 to 3 active flavors, the equation of state changes
so that $T_0$ actually declines.

Regardless of the parameters employed, it is apparent that
minijet production yields substantially larger initial temperatures than
previously predicted (for example, in Ref.~\cite{Satz91})
and thus enhances the possibility of QGP production in
thermal equilibrium. Only the extent of this enhancement
is uncertain. For large enough $p_0$ and/or extensive nuclear shadowing, the
energy density and temperature might be reduced sufficiently that
the plasma would not be fully equilibrated, although 
a QGP might still be formed. In any case, it is important to examine
the proposed signatures of plasma formation for the increased
initial energy density and temperature obtained after including
the effects of minijet production.

One of the proposed signatures of the QCD phase transition is the suppression
of quarkonium production, particularly of the $J/\psi$ \cite{MS,KMS}.  In
fact, $J/\psi$ and $\psi^\prime$ suppression have been observed in
nucleus-nucleus collisions at the CERN SPS with $\sqrt{s} = 19.4$ GeV
\cite{NA38,NA50}.  In a QGP, the suppression
occurs due to the shielding of the $c \overline c$ binding potential by color
screening, leading to the breakup of the resonance.  The $c \overline c$
($J/\psi$, $\psi'$, $\cdots$) and $b
\overline b$ ($\Upsilon$, $\Upsilon'$, $\cdots$)
resonances have smaller radii than light-quark hadrons and
therefore higher temperatures are needed to disassociate these
quarkonium states.
At current energies, the situation for the $J/\psi$ is rather ambiguous because
the bound state can also break up through interactions with nucleons and
comoving hadrons---QGP production is not a unique explanation of $J/\psi$
suppression \cite{GV,GSTV,DH,GV2}.  At LHC energies there 
will continue to be ambiguity in this regard, and, in addition,
$J/\psi$ and $\psi^\prime$ detection efficiencies may not be large \cite{CMS}.

In contrast, the $\Upsilon$ 
resonances are sufficiently tightly bound that it is more difficult 
to break them up as a consequence of ordinary multiple interactions. 
In addition, they can be detected in the $\mupmum$ decay
mode with relatively high efficiency at the LHC (\eg 33\% vs.\ 6\% for
$J/\psi$ in the CMS detector \cite{CMS}).
Further, this detection efficiency derives entirely from
the central region ($|\eta|<2.4$ in the CMS detector \cite{CMS}), 
where the predictions for the relative 
suppression of the members of the $\Upsilon$
family in the presence of a gluon plasma can be most reliably tested
(given that the baryon density is predicted
to be low and should not play a significant role).
However, it was previously assumed that the $\Upsilon$ would not 
be suppressed by QGP production \cite{KMS,KS},
since the $\Upsilon$ is much smaller
than the $c \overline c$ and other $b \overline b$ resonances (such
as the $\Upsilon'$ and $\chi_b$ states), implying that a much higher
temperature than expected from earlier estimates of $T_0$ 
is needed to dissociate the $\Upsilon$ \cite{KMS}.    
In view of the high initial temperature 
of a gluon plasma, $T_0 \approx 1$ GeV, when minijets are included,
a reexamination of this assumption is in order.  If $\Upsilon$ suppression by
a minijet plasma is predicted, the $\Upsilon$ may provide a cleaner
plasma signature than the $J/\psi$ because of its weaker interactions with
matter.  In high energy heavy-ion collisions, the $\Upsilon$
rate should be large enough to test the predictions given here
and help clarify the initial state of the system.

The paper is organized as follows.  In the next section, we review the model
of quarkonium suppression in a QGP and introduce two different scenarios for
quarkonium breakup at finite temperature.  Section 3 presents the quarkonium
production model we adopt and includes a  discussion of the expected $\Upsilon$
production rate. We then provide a detailed examination of $\Upsilon$ family
suppression in a minijet plasma as a function of $p_T$ and discuss the
sensitivity of the results to the initial conditions.  In section 5, we
elucidate the steps necessary to extract information on the state of the
system.  We particularly focus on the $\Upsilon'/\Upsilon$ ratio and discuss
the effect of cascade decays from higher states on the signal.
We also propose using the
$Z$ boson as a reference process against which $\Upsilon$ suppression can be
tested.  In summary, we show that the $\Upsilon$ family may provide
an extremely powerful tool for diagnosing the existence and nature
of the quark-gluon plasma.

\section{Quarkonia in a quark-gluon plasma}

\indent\indent
We follow the arguments of Ref.\ \cite{KMS} to discuss the finite temperature
properties of quarkonium.  The interquark potential for nonrelativistically
bound quarkonium at zero temperature is
\begin{eqnarray} 
V(r,0) = \sigma r - \frac{\alpha}{r} \, \, , 
\label{vform}
\end{eqnarray}
where $r$ is the separation between the $Q$ and $ \overline Q$.  The
parameters are $\sigma = 0.192$ GeV$^2$, $\alpha = 0.471$, $m_c = 1.32$ GeV and
$m_b = 4.746$ GeV \cite{KMS}.  
This potential is modified at finite temperatures due to
color screening, becoming 
\begin{eqnarray} 
V(r,T) = 
\frac{\sigma}{\mu(T)} (1 - e^{-\mu(T) r}) - \frac{\alpha}{r} e^{-\mu(T) r}
\, \, .
\label{vrt}
\end{eqnarray} 
The screening mass, $\mu(T)$, is an increasing 
function of temperature.  When $\mu(T=0) \rightarrow 0$, eq.~(\ref{vform}) is
recovered.  At finite temperature, when $r \rightarrow 0$, the $1/r$ behavior
is dominant while as $r \rightarrow \infty$ one finds that the range of the
potential decreases with $\mu(T)$, making the binding less effective at finite
temperature.  Semiclassically the energy is 
\begin{eqnarray} 
E(r,T) = 2m_Q + \frac{c}{m_Q r^2} + V(r,T) \, \, , 
\label{ert}
\end{eqnarray} 
where $\langle p^2
\rangle \langle r^2 \rangle = c \approx {\cal O}(1)$ \cite{KMS}.  Minimizing
$E(r,T)$ gives the radius of the bound state at each $T$.  For $\mu(T)$
above a certain critical value, $\mu_D$,
there is no longer a minimum and the screening 
has become strong enough for binding to be
impossible and the resonance no longer forms in the
plasma.  The temperature $T_D$ at which this happens is that
for which $\mu(T_D)=\mu_D$.

The values of
$\mu_D$ for the bottonium states are given in Table~\ref{tablei},
along with the bottonium mass at $\mu=\mu_D$, $M(\mu_D)$.
Also shown are the results for the charmonium states.
We note that $\mu_D$ and $M(\mu_D)$ are independent
of $T_D$, but that $T_D$ depends upon the functional form of $\mu(T)$.  
Table~\ref{tablei} shows that the
$\Upsilon$ mass is actually larger at breakup than at $T=0$ because the $1/r$
contribution to the potential is dominant, whereas
the $\Upsilon'$ and $\chi_b(1P)$ masses are smaller due to the
fact that the linear term is more important for these resonances.

\begin{table}[htb]
\begin{center}
\begin{tabular}{|c|c|c|c|c|c|c|} \hline
& $\Upsilon$ & $\Upsilon'$ & $\chi_b(1P)$ & $J/\psi$ & $\psi'$ & $\chi_c(1P)$
\\ \hline
$M$ (GeV) & 9.445 & 10.004 & 9.897 & 3.07 & 3.698 & 3.5 \\ \hline
$r$ (fm) & 0.226 & 0.509 & 0.408 & 0.453 & 0.875 & 0.696 \\ \hline
$\tau_F$ (fm) & 0.76 & 1.9 & 2.6 & 0.89 & 1.5 & 2.0 \\ \hline
$M(\mu_D)$ (GeV) & 9.615 & 9.778 & 9.829 & 2.915 & 3.177 & 3.198 \\ \hline
$\mu_D$ (GeV) & 1.565 & 0.671 & 0.558 & 0.699 & 0.357 & 0.342 \\ \hline
\end{tabular}
\end{center}
\caption[]{Properties of the quarkonium states both at zero temperature and at
$T = T_D$ (the breakup temperature for each bound state), taken from Ref.\
\cite{KMS}.  The masses at $T=0$
are from the solution to the Schrodinger equation with the potential of 
eq.~(\ref{vform}).
Note that $r$ is the distance between the $Q$ and $\overline Q$ and
$\tau_F$ is the formation time of the bound state.  $M(\mu_D)$ and $\mu_D$ are
the mass at breakup and the screening mass at $T_D$; both
are independent of the actual value of $T_D$ as determined
by the functional form of $\mu(T)$.}
\label{tablei}
\end{table}

\begin{figure}[htb]
\let\normalsize=\captsize   
\begin{center}
\centerline{\psfig{file=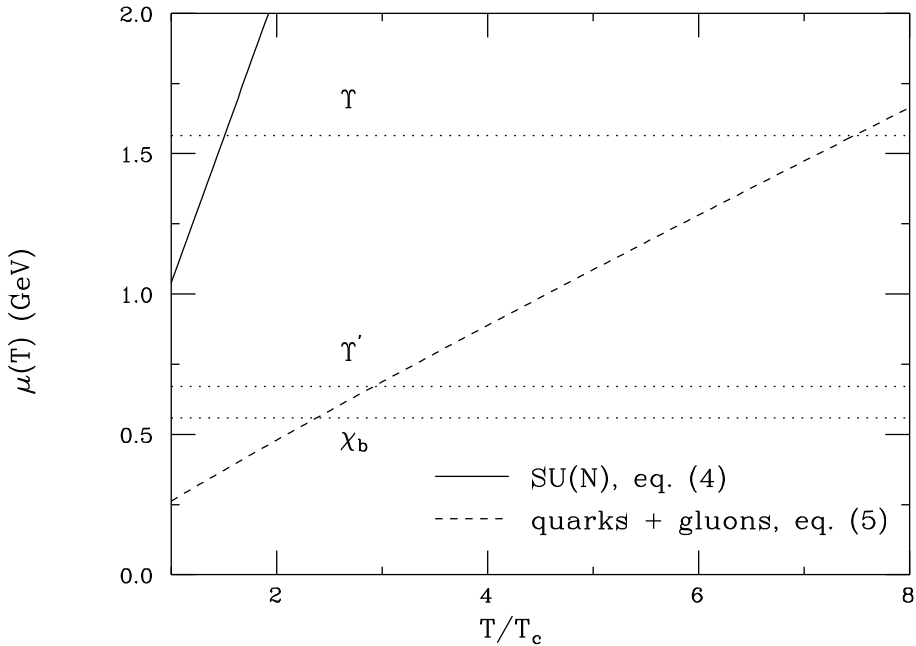,width=12.2cm}}
\begin{minipage}{12.5cm}       
\caption{
The screening mass as a function of temperature for the scenarios
of eqs.~(\protect\ref{mut1}) (solid) and (\protect\ref{mut2}) (dashed).  
The values of $\mu_D$ at breakup for the $\Upsilon$,
$\Upsilon'$, and $\chi_b$ are also indicated by the dotted lines.}
\label{figurei}
\end{minipage}
\end{center}
\end{figure}

It is necessary to know the dependence of the screening mass on
temperature before the breakup temperature
$T_D$ can be determined.  Since the behavior of $\mu(T)$ is
unknown for $T \gg T_c$, we consider two scenarios which may be
expected to bracket the realistic situation.  
\begin{itemize}
\item
The first is a
parameterization based on SU($N$) lattice simulations
\cite{Petersson}, 
\begin{eqnarray} 
\frac{\mu(T)}{T_c} \simeq 4 \frac{T}{T_c} 
\label{mut1}
\end{eqnarray} 
which results in the lowest values of $T_D$.  Although this parameterization
of $\mu(T)$ is independent
of $T_c$, we will present our results as a function of the ratio $T/T_c$
taking $T_c = 260$ MeV, consistent with SU(3) lattice calculations.  For this
value of $T_c$, the $\Upsilon$ breaks up at $T_D/T_c \approx 1.5$.  In Ref.\
\cite{KS}, this parameterization was used with $T_c = 150$ MeV, leading to the
estimate $T_D/T_c \sim 2.6$ for the $\Upsilon$.  
\item
The second is an estimate from perturbation theory \cite{GPY},
\begin{eqnarray} 
\frac{\mu(T)}{T_c} = \sqrt{1 +
\frac{n_f}{6}} g \left(\frac{T}{T_c} \right) \frac{T}{T_c} \, \, ,
\label{mut2}
\end{eqnarray} 
where the temperature-dependent running coupling constant is
\begin{eqnarray} 
g^2 \left(\frac{T}{T_c} \right) = \frac{24\pi^2}{(33-2n_f) \ln
[(19T_c/\Lambda_{\rm \overline{MS}})(T/T_c)]} \, \, . 
\label{g2t}
\end{eqnarray} 
In SU(3) gauge theory, $T_c/\Lambda_{\rm \overline{MS}} = 1.78 \pm 0.03$
\cite{KMS}.  In this case, $T_D$ can be quite high and depends
strongly on $n_f$ and $T_c$.  We present results using eq.~(\ref{mut2}) with
$n_f = 3$ and $T_c = 150$ MeV.  
\end{itemize}
We shall refer to these two distinct choices for $\mu(T)$ as
the SU($N$) and 3-Flavor QGP's, respectively.
The temperature dependence for both estimates of $\mu(T)$ is shown in 
Fig.~\ref{figurei} with $\mu_D$ indicated 
for $\Upsilon$, $\Upsilon'$ and $\chi_b$. 
The values of $T_D$ are given in Table~\ref{tdtable} for both cases.
Even in the first case, the early estimates
of initial temperatures of order $T_0\sim 260$ MeV for 
the LHC implied that $\Upsilon$ suppression would be unlikely \cite{Satz91},
and that even $\Upsilon'$ and $\chi_b$ suppression would not be certain.

\begin{table}[htb]
{\small
\begin{center}
\begin{tabular}{|c|c|c|} \hline
\multicolumn{3}{|c|}{$T_D$ (MeV)} \\ \hline
& SU($N$) only, eq.~(\ref{mut1})  &
 gluons and quarks, eq.~(\ref{mut2}) \\ \hline
$\Upsilon$ & 391 & 1125 \\ \hline
$\Upsilon'$ & 260 & 447 \\ \hline
$\chi_b$ & 260 & 368 \\ \hline
$J/\psi$ &260 & 459 \\ \hline
$\psi'$ & 260 & 211.5 \\ \hline
$\chi_c$ & 260 & 204 \\ \hline
\end{tabular}
\end{center}
}
\caption[]{The values of $T_D$ (in MeV) for the two
choices of $\mu(T)$, eq.~(\ref{mut1}) and eq.~(\ref{mut2}).
We employ $T_c=260$ MeV for the SU$(N)$ case and
$T_c=150$ MeV in the case of eq.~(\ref{mut2}).  The lower $T_c$ used in the
3-Flavor case results in the lower $T_D$ values for the $\psi'$ and $\chi_c$.}
\label{tdtable}
\end{table}

In contrast, the higher `minijet' value of $T_0\sim 1.14\gev$, combined with
the results of Fig.~\ref{figurei} and the $T_D$ values
given in Table~\ref{tdtable} indicate that,
when the SU$(N)$ parameterization 
is used, the $\Upsilon$ may easily be suppressed
in the minijet gluon plasma at LHC energies.  
However, the 3-Flavor estimate 
leads to much higher breakup temperatures for the Upsilon states 
and the distinct possibility that the $\Upsilon$ will not be 
suppressed\footnote{Note that $\Upsilon$ production from 
$\Upsilon'$ and $\chi_b$
decays will be suppressed if the  $\Upsilon'$ and $\chi_b$ states are
suppressed.} even though the $\Upsilon'$ and $\chi_b$ could be suppressed. 
In either case, the suppression pattern depends on the
space-time evolution of the plasma. Depending upon circumstances,
this dependence could be either an impediment to the extraction
of the physics of the QGP or a very powerful tool.
In most scenarios, much can be learned about the QGP 
by simply examining the ratio \cite{KS} 
of production rates for higher $b\anti b$ 
resonances to the production rate for the $\Upsilon$. However,
for a quantitative determination of QGP suppression on an absolute
scale, a benchmark such as $Z$ production,
to which $\Upsilon$ production cross sections may be compared, is needed. 

\section{Cross sections, rates, and the color evaporation model}

\indent\indent
Before turning to the details of $\Upsilon$ suppression, let us
determine whether or not the $\Upsilon$ 
production rate is large enough for
any suppression to be observable.  An estimate of the $\Upsilon$ yield is not
straightforward since the formation of bound states from the predominantly 
color octet $Q \overline Q$ pairs is nonperturbative and therefore 
model-dependent. 

The color evaporation model \cite{Schuler}
has often been applied to
the production of quarkonium states below the open
charm/bottom thresholds.  The $Q \overline Q$ pair neutralizes its color by 
interaction with the collision-induced color field---``color evaporation".
The $Q$ and the $\overline Q$ either combine with light
quarks to produce heavy-flavored hadrons or bind with each other 
in a quarkonium state.  The additional energy needed to produce
heavy-flavored hadrons is obtained nonperturbatively from the
color field in the interaction region.
Depending on $m_Q$, the yield of all quarkonium states
may be only a small fraction of 
the total $Q\anti Q$ cross section below the $\sqrt{\hat s}=2m_H$ threshold,
where $m_H$ is the mass of the lightest heavy-flavored meson.
The total heavy quark cross section below this $H \overline H$ 
threshold, $\tilde{\sigma}$, is calculated at leading order by
integrating over the $Q \overline Q$ pair mass from $2m_Q$ to
$2m_H$,
\begin{eqnarray}
\tilde{\sigma}(s) = \sum_{i,j} \int_{4m_Q^2}^{4m_H^2} d\hat s \int dx_1 
dx_2~f_{i/p}(x_1)~f_{j/p}(x_2)~ \sigma_{ij}(\hat s)~\delta(\hat s-x_1x_2s)\, \, ,
\label{sigtil}
\end{eqnarray} 
where $ij = q \overline q$ or $gg$ and $\sigma_{ij}(\hat s)$ is the
$ij\to Q\anti Q$ subprocess cross section.
The color evaporation model was
taken to next-to-leading order (NLO) using the exclusive $Q \overline
Q$ hadroproduction calculation
\cite{MNR} to obtain the energy, $x_F$, and $p_T$-dependence
of quarkonium production \cite{Gavai,SchulV}.

The division of $\tilde{\sigma}$ into heavy flavored hadrons vs.
quarkonium and the relative quarkonium production rates are both
a matter of assumption in the color evaporation model.
For the model to have any predictive power,
the relative quarkonium production rates should be independent of
projectile, target, and energy.  Experiment suggests that this is 
true for the charmonium
ratios $\chi_c/ \psi$ and $\psi^\prime/\psi$ over a broad energy
range \cite{Teva,Anton1,Anton2,Ronceux}.
The available bottonium data follows this trend:
$\Upsilon'/\Upsilon = 0.53 \pm 0.13$ and $\Upsilon''/\Upsilon = 0.17 \pm 0.06$
for proton beams at 400 \cite{Ueno} and 800 GeV \cite{Yoshi,Moreno}.  These
results are consistent with the production of $\Upsilon$ states in $p \overline
p$ collisions at the Tevatron \cite{Eggert2}.  The color
evaporation model also reproduces the energy dependence of charmonium
and bottonium production over a wide range of energy as well as most 
of the $x_F$
dependence of the charmonium states\footnote{At high $x_F$, additional
production from other mechanisms such as intrinsic heavy quarks \cite{Stan} 
may be important.}, provided the fraction of quarkonium vs. $H\anti H$
states is assumed to be independent of all kinematic variables.

With these assumptions and our knowledge of the $\Upsilon'/\Upsilon$
and $\Upsilon''/\Upsilon$ ratios at low energy,
the normalization of each quarkonium state can be fixed empirically from
data, allowing predictions of the production cross sections at LHC energies. 
The procedure we have followed is described below.
\begin{itemize}
\item
As a first step, we assess the accuracy of the model for 
describing existing $pp$/$p\anti p$ data.
Fixed target $\Upsilon$ data has been generally given as the sum
of $\Upsilon$, $\Upsilon'$, and $\Upsilon''$ production.  From this data, 
one can determine $B(d\sigma/dy)_{y=0}$ where $B$ is an effective dilepton
branching ratio reflecting the production yield through the
$\Upsilon$, $\Upsilon'$, and $\Upsilon''$ resonances. 
As shown in Fig.~\ref{figureii}, taken from Ref.\ \cite{Gavai},
a good fit to the data, up to and including ISR energies
\cite{Ueno,Yoshi,Moreno,fixt,ISR}, is obtained with
\begin{eqnarray} 
B\left({d\sigma (s)\over dy}\right)_{y=0}
= 1.6~\times 10^{-3}~\left( {d\tilde{\sigma}(s)\over dy}
\right)_{y=0}, 
\label{bsigy}
\end{eqnarray}
where $d\tilde\sigma/dy$ is computed
using the MRS D$-^\prime$ \cite{MRS} or GRV HO 
\cite{GRV} PDFs, taking
$m_b=4.75$ GeV and the renormalization and factorization scales set to
$\mu=m_{T, b \overline b} =
\sqrt{m_b^2 + (p_{T,b}^2 + p_{T, \overline b}^2)/2}$.
The high energy data from UA1 \cite{Eggert1} and CDF \cite{Eggert2}
are also shown in Fig.~\ref{figureii} and agree very well
with the energy dependence of the color evaporation model,
as obtained from eq.~(\ref{sigtil}).
Both sets of PDFs
give compatible predictions at the LHC energy.
The predicted cross section 
is up to a factor of 20 larger at 5.5 TeV
than that given by the empirical parameterization,
$d \sigma/dy|_{y=0} \propto {\rm exp}(-14.7 M_\Upsilon/\sqrt s)$ 
\cite{Craigie,Ramona}, labeled CR in Fig.~\ref{figureii}.

\begin{figure}[htb]
\let\normalsize=\captsize   
\begin{center}
\centerline{\psfig{file=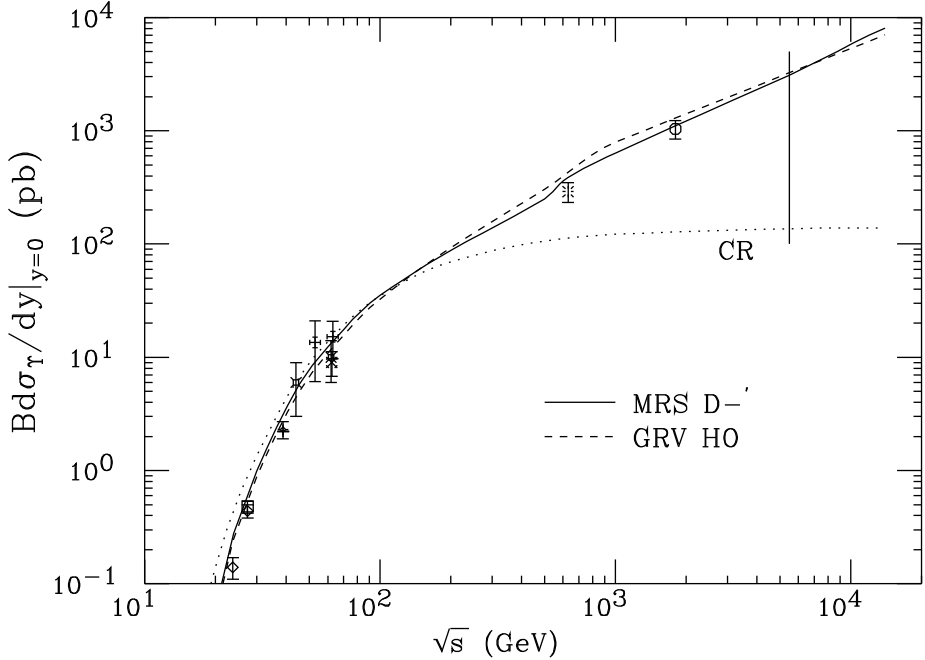,width=12.2cm}}
\begin{minipage}{12.5cm}       
\caption[]{
We show $Bd\sigma/dy|_{y=0}$ for $\Upsilon + \Upsilon' +
\Upsilon''$ in $pp$ collisions, 
as indicated in eq.~(\protect\ref{bsigy}), for the MRS D$-^\prime$ 
\protect\cite{MRS}
(solid) and GRV HO \protect\cite{GRV} (dashed)
parton densities.  Also shown is the empirical Craigie parameterization 
\protect\cite{Craigie,Ramona} (dotted).  The data are taken from
\protect\cite{Ueno,Yoshi,Eggert2,fixt,ISR,Eggert1}.  
The vertical solid line indicates the maximum LHC Pb$+$Pb energy.
This figure was first presented in Ref.\ \cite{Gavai}.}
\label{figureii}
\end{minipage}
\end{center}
\end{figure}

\item The second step is to
extract the $\Upsilon$, $\Upsilon'$ and $\Upsilon''$
components separately using
the $\Upsilon'/\Upsilon$ and $\Upsilon''/\Upsilon$ ratios quoted earlier,
and the known branching ratios 
$B_{\Upsilon_i} \equiv B(\Upsilon_i \rightarrow
\mu^+ \mu^-)$ where $\Upsilon_i$ represents the individual $\Upsilon$ states.
If we write $d\sigma_{\Upsilon_i}/dy|_{y=0} \equiv
 f_{\Upsilon_i} d\tilde{\sigma}/dy|_{y=0}$ for the 
$\Upsilon_i$ cross sections, then from eq.~(\ref{bsigy})
\begin{eqnarray}
f_{\Upsilon}B_{\Upsilon} + f_{\Upsilon'}B_{\Upsilon'} + 
f_{\Upsilon''}B_{\Upsilon''} = B=0.0016 \,.
\label{fbsum}
\end{eqnarray}
Using $f_{\Upsilon'}/f_{\Upsilon}=0.53$ and 
$f_{\Upsilon''}/f_{\Upsilon}=0.17$, as quoted earlier,
and $B_{\Upsilon_i}$ from the Particle Data Book \cite{PDG}, we find
\cite{SchulV} 
\begin{equation}
f_{\Upsilon} = 0.044\,,\quad f_{\Upsilon'} = 0.023\,,\quad f_{\Upsilon''} = 
0.0074\,.
\label{thefs}
\end{equation}

\item Thirdly, we must separate direct from indirect production. 
The measured $\Upsilon_i$ production cross sections, or equivalently the 
$f_{\Upsilon_i}$, are only effective values which reflect both direct 
production and chain decays of higher mass states. 
Isolation of the direct production cross sections
for each $\Upsilon_i$ requires
detection of the radiated photons associated with chain decays.  Data
which allows this separation is not currently available. As discussed
later, the extent to which the required photon detection
will be possible at the LHC is also uncertain. We shall see that
this is a very important issue. For now, we make the assumptions:
\begin{equation}
f_{\chi_{bi}(1P)}^d=f_{\Upsilon'}\,;\quad f^d_{\chi_{bi}(2P)}=f_{\Upsilon''}\,,
\label{fchis}
\end{equation}
for {\it each} $i$, where $i=0,1,2$ labels the $\chi_b$ states in 
PDG notation \cite{PDG},
and the $d$ superscript indicates the $f$ for {\it direct} production.
With this and the summed branching ratios \cite{PDG}
\begin{eqnarray*}
&&\sum_{i=0,1,2}B(\chi_{bi}(1P) \to \Upsilon \gamma) \sim 0.63\,, \\
&&\sum_{i=0,1,2}B(\chi_{bi}(2P) \to \Upsilon \gamma) \sim 0.16\,, \\
&&\sum_{i=0,1,2}B(\chi_{bi}(2P) \to \Upsilon' \gamma) \sim 0.42\,,
\label{sumbrs}
\end{eqnarray*}
along with $B(\Upsilon'\to\Upsilon X)\sim 0.27$ and $\Upsilon''$ decays to
$\Upsilon'$, $\Upsilon$ and $\chi_{bi}(2P)$, we can compute the
$f^d$'s for direct production of all the $\Upsilon_i$.
One finds that about 0.014 of the total 
$f_{\Upsilon}=0.044$ is due to $\chi_{bi}(1P)$ decays.
This would be similar to what is found in the analogous $\psi$, $\chi_c$
situation. Similarly, $\sim 0.001$
of $f_{\Upsilon}$ and $\sim 0.003$ of $f_{\Upsilon'}$
would be due to $\chi_{bi}(2P)$ decays.
Also important are $\Upsilon'\to\Upsilon X$
decays; $f_{\Upsilon'}$ [eq.~(\ref{thefs})] implies that an additional
0.006 of $f_{\Upsilon}$ would be indirect.  The contributions from
chain decays of the $\Upsilon''$ are small. Also,
there are no contributions to the $\Upsilon''$ 
rate coming from higher states that are known to be significant.
Altogether, we would have:
\begin{eqnarray}
&f^d_{\Upsilon}\sim 0.023\,,~f^d_{\Upsilon'}\sim 0.020\,,~
f^d_{\Upsilon''}\sim 0.0074\,, & \nonumber \\
&f^d_{\chi_{bi}(1P)}\sim 0.023\,,~f^d_{\chi_{bi}(2P)}\sim 0.0074\,, &
\label{fds}
\end{eqnarray}
where $i=0,1,2$ labels the different $\chi_{bi}(1P,2P)$ states.
Note that only about half of $f_{\Upsilon}$ 
is due to direct $\Upsilon$ production. 
The results in eq.~(\ref{fds}) imply 
that about 14\% of $\tilde{\sigma}$ goes into bottonium state production.  
\end{itemize}

In Table~\ref{tablerates}, we give the corresponding normalized 
direct production cross sections in $pp$ collisions, $f^d\tilde\sigma$, 
for each state with $f^d$ from
eq.~(\ref{fds}) and $\tilde\sigma$ computed
using the MRS  D$-^\prime$ parton densities at 5.5 TeV.
The cross sections 
given in Table~\ref{tablerates} are integrated over all rapidity. 
At this energy, the results with the GRV HO 
distributions are about 20\% smaller.

\begin{table}[htb]
\begin{center}
\begin{tabular}{|c|c|c|c|c|c|} \hline
& $\Upsilon$ & $\Upsilon'$ & $\Upsilon''$  & $\chi_b(1P)$ & $\chi_b(2P)$ 
\\ \hline
$f^d\tilde\sigma$ (nb) & 426 & 233 & 137 & 426 & 137 \\ \hline
$N^d({\rm PbPb})$ & $2.02 \times 10^6$ & $1.75 \times 10^6$ & 
$6.47 \times 10^5$ & $2.02\times 10^6$ & $6.47\times 10^5$ \\ \hline
$N^d_{\mu \mu}({\rm PbPb})$ &  $4.99 \times 10^4$ & $2.30 \times 10^4$ 
& $1.17 \times 10^4$ & - & -\\ \hline
\end{tabular}
\end{center}
\caption[]{The normalized cross sections,
$f^d\tilde{\sigma}$, for directly produced bottonium states in $pp$
collisions at 5.5 TeV, using
the $f^d$'s of eq.~(\ref{fds}) and the prediction of
eq.~(\ref{sigtil}) that $\tilde{\sigma} =
17.35$ $\mu$b \cite{Gavai}, where
the MRS D$-^\prime$ parton densities are employed with $m_b = 4.75$
GeV and $\mu=m_{T,b\overline b}$. Also given
is the number $N^d$ of each type of bottonium state produced 
directly in Pb+Pb collisions
at the LHC, calculated using eq.~(\ref{upsrate})
with $\sigma_{NN}=60$ mb, $T_{\rm PbPb}({\bf 0})=32$ mb$^{-1}$ and
$L_{\rm int}^{\rm PbPb}=2.59$ nb$^{-1}$. 
For $\Upsilon,\Upsilon',\Upsilon''$ we give the corresponding
number of $\mu^+\mu^-$ pairs from decay of directly produced states.
}
\label{tablerates}
\end{table}

We would like to comment on the differences between the color evaporation
model used here and other models of quarkonium production.
The color singlet model \cite{BR}, in which bound states with the 
right quantum numbers are directly produced, has been shown to
be insufficient to reproduce the Tevatron data on $J/\psi$, $\psi'$, 
and $\Upsilon$ production \cite{Teva,Eggert2}.  The low-$p_T$ $\Upsilon$
production data from the Tevatron has recently been explained using a model
of color octet production involving corrections to the bound state
wavefunctions at higher order in the relative velocity of the $Q \overline Q$
where $\alpha_s \sim v$ for the bound state \cite{Cho}.  We would like to 
point out that the $p_T$-dependence of quarkonium production is quite similar
to the $p_T$-dependence of unbound $Q \overline Q$ pairs \cite{MNR}.  
This same dependence
also appears in the color octet model since the octet mechanism makes other
$^3S_1$ production channels available.  Thus the color octet model 
is quite similar in spirit to the color evaporation model considered here.
However, at larger $p_T$, $p_T > m_\Upsilon$, fragmentation processes such as
$g \rightarrow \Upsilon$ and $b \rightarrow \Upsilon$ \cite{frag}
should eventually dominate the
production processes considered in the color octet
model.  It has been recently shown that the Tevatron charmonium and bottonium 
$p_T$ data is also in good agreement with the color evaporation model at
NLO, as described in \cite{SchulV}.  In the color evaporation
picture, $gg$ scattering followed by the splitting
$g \rightarrow b \overline b$ incorporated at NLO
is similar to models of $g \rightarrow \Upsilon$
fragmentation \cite{frag}.  By including
this splitting, the color evaporation model provides a good
description of the quarkonium $p_T$ distributions.

We can now employ the results for $f^d\title\sigma$
given in Table~\ref{tablerates}
to predict rates for direct production of
bottonium states in Pb+Pb collisions at the LHC.
For central collisions, the expected rates are given by
\begin{eqnarray} N^d = \sigma_{NN}
T_{\rm PbPb}({\bf 0}) f^d\tilde\sigma L_{\rm int}^{\rm PbPb} \, \, ,
\label{upsrate}
\end{eqnarray} 
where $\sigma_{NN} T_{\rm PbPb}({\bf 0})$ is the number of
central Pb+Pb collisions\footnote{We use $\sigma_{NN} \approx 60$ mb, assuming
that the high energy rise in $\sigma_{p \overline p}$ will also hold for
$\sigma_{NN}$, and $T_{\rm PbPb}({\bf 0})\simeq 32$/mb.}
and $f^d\tilde\sigma$, with $f^d$ taken from eq.~(\ref{fds}), is
the cross section for direct production
of a given bottonium state in $pp$ collisions. In one month of running the
integrated luminosity for lead beams is expected to be
$L_{\rm int}^{\rm PbPb} = 2.59/$nb.
Typical rates are given in Table~\ref{tablerates}; they are on the order of
10$^6$ for $\Upsilon$ and $\Upsilon'$. Approximately 10-15\% of the 
cross section is within $|\eta|\leq1$.
The number of muon pairs from the
$\Upsilon$, $\Upsilon'$ and $\Upsilon''$ decays, 
found by multiplying the total number of
$\Upsilon_i$ directly produced, $N^d$, by the
appropriate branching ratio, is also given in Table~\ref{tablerates}.
These rates suggest that production and
suppression of these states should be
measurable by CMS in the very clean $\mupmum$ final state decay mode.  
The rapidity distributions are rather broad and nearly
constant out to $y\simeq 4$ with the MRS D$-^\prime$ 
parton distributions.  The GRV HO results exhibit a somewhat narrower rapidity
plateau \cite{Gavai}.

The rates of Table~\ref{tablerates} are those obtained by simply multiplying
the $pp$ cross section by the expected number of central Pb+Pb collisions. 
This procedure does not include any nuclear 
effects (\eg shadowing or absorption) on $\Upsilon$ production.  
However, significant nuclear effects have been observed in fixed-target 
interactions. 
The E772 collaboration has measured a less than linear $A$ dependence for
$\Upsilon$ production at $\sqrt{s} = 38.8$ GeV, $\sigma_{pA}^\Upsilon =
\sigma_{pp}^\Upsilon A^{\alpha}$ where $\alpha = 0.96$ \cite{E772}.  
If no nuclear effects were 
present, the production cross section would grow as $A^1$.  
Note that $\alpha=0.96$ implies a 20\% reduction in the $p$Pb 
production cross section relative to the $A^1$ prediction.  
The two possible sources of this reduction in $pA$
collisions are shadowing and absorption.
\begin{itemize}
\item If the reduction is due to shadowing, the observed greater
suppression (smaller $\alpha$) of $J/\psi$ compared to $\Upsilon$ implies
that the shadowing of the gluon PDFs is greater at smaller $x$ values.
This would then imply that, for the $\Upsilon$ (or any other given
onium resonance), shadowing would
be greater at higher energies due to the lower $x$ region probed. 
In heavy-ion collisions at LHC energies the shadowing mechanism 
could reduce $\sigma_{\rm PbPb}^\Upsilon$ by 25-40\%
relative to the result obtained assuming proportionality to $A^2$
\cite{Eskolash}. The larger shadowing effect expected for Pb+Pb collisions
compared to $p$Pb collisions comes from 
the convolution of two nuclei with modified parton densities. 
To the extent that shadowing in central collisions is the
same as for minimum bias collisions (which include peripheral
as well as central collisions), the same 25-40\% reduction
should be applied to the numbers of Table~\ref{tablerates}.
However, nucleons near the nuclear surface might experience a smaller
shadowing effect than those in the center of the nucleus.
\item
On the other hand, the deviation of the $A$ dependence from
unity in fixed-target interactions has also been attributed to absorption
by nucleons; interactions with comoving secondaries are possible although
rare in $pA$ collisions. 
In this approach, $\alpha$ is related to the
the effective onium absorption cross section in nuclear matter
by $1-\alpha \approx 9\sigma_{\rm
abs}/16\pi r_0^2$, assuming a step function nuclear shape.
The apparent $\Upsilon$ absorption cross section 
is 40-50\% less than the $\psi$ absorption cross section, suggesting 
that the
$\Upsilon$ suffers fewer final-state interactions than the $J/\psi$
(possibly because of its smaller size).
The onium-nucleon absorption cross section  has been calculated on the basis
of the operator-product expansion \cite{Pes,BhP} and has been shown to quickly
attain its asymptotic value \cite{KSSZ}. 
Absorption by nucleons has been shown
to be less effective in peripheral relative to central nuclear collisions, see
{\it e.g.}\ \cite{GSTV,GV2}.   Therefore nucleon interactions with the
$J/\psi$ may be insufficient to account for $J/\psi$ measurements in heavy-ion
interactions at current energies \cite{NA38,NA50}.
Secondary particle production increases in 
central relative to peripheral collisions, increasing the influence of
comovers in central collisions as well as in nuclear collisions compared to
$pA$ interactions \cite{GV2}.  The onium-comover cross section can be
estimated from the onium-nucleon absorption cross section by the additive
quark model assuming the comovers are mesons \cite{Gena}. 
\end{itemize}
A mixture of shadowing and absorption is most likely.
If the value of $\alpha$ depends on the Upsilon state,
absorption is
important: $\sigma_{\rm abs}$ could increase with the 
size of the Upsilon state \cite{PH}.  If shadowing were dominant, $\alpha$
would remain essentially unchanged since the $x$ values probed by the different
Upsilon states are very nearly equal.  In fact, the E772 collaboration 
observed no difference in the $J/\psi$ and $\psi'$ $A$ dependence or in the
$\Upsilon$ and $\Upsilon'$ $A$ dependence within their uncertainties
\cite{E772}.  Additionally the $\psi'/\psi$ ratio appears to be independent of
$A$ \cite{Gavai} but the data is such that a size-dependence of the absorption
cross section cannot be ruled out.  At the E772 energy, $\sqrt{s} = 38.8$ GeV,
the target momentum fractions probed are $x \leq 0.08$ for the charmonium
states, just in the shadowing region, while for the $\Upsilon$ states, $x \leq
0.25$, in the EMC or antishadowing regions.  At the lower energies of other
$J/\psi$ measurements, shadowing should be less effective than absorption
\cite{VBH}.  The mass scale in the PDF is also important if the shadowing is
$Q^2$ dependent \cite{LV}.  In $Q^2$-dependent shadowing models, the gluon
distribution evolves fastest with $Q^2$; the valence and sea quark evolution
is rather weak for $2<Q^2<10$ GeV$^2$ \cite{Eskolash,LV}.  We note that nuclear
deep-inelastic scattering data do not exhibit strong $Q^2$ dependence for
$Q^2 \leq 100$ GeV$^2$ \cite{Arn}, in agreement with these models, but it is
more difficult to probe the $Q^2$ dependence of gluon shadowing.
Precision $pA$ data is needed to sort out these
effects, particularly at higher than current energies since at RHIC and LHC
we will be in a new regime where $x$ is small and $Q^2$ is relatively large.  
Measurements of dilepton production will prove helpful, as we discuss later, 
since the dilepton continuum could 
be influenced by shadowing but not by absorption.

The $p_T$-dependence of the suppression from shadowing/absorption is especially
crucial for out later considerations.
Since $pA$ studies are not possible at
the LHC, the exact rapidity and $p_T$-dependence of shadowing/absorption 
suppression at the low $x$ values probed at $5.5\tev$ will be 
imprecisely known. Thus, it is fortunate
that the $p_T$-dependence of such suppression should be
much weaker than that from QGP suppression (as we shall discuss below).
As noted above, if shadowing is dominant, then suppression
from this purely nuclear source should also be essentially
independent of the Upsilon state.   As our ensuing
discussion will show, this would be a very important advantage.
In contrast, if absorption is dominant the magnitude of the
suppression could depend on the Upsilon state.
Finally, we note that any
broadening of the $p_T$-dependence due to initial-state interactions in the
nuclear target \cite{GG} may be expected to be negligible.
Thus, we believe that any alteration of the $p_T$-spectra for 
the various members of the $\Upsilon$ family due to QGP effects will stand out
clearly. Even more crucially, 
in the absence of QGP suppression the ratios of rates for
different $\Upsilon$ family members should exhibit $p_T$
dependence that is precisely that predicted
without including shadowing or absorption.\footnote{If absorption
is significant and $\sigma_{\rm abs}$ depends upon the Upsilon state,
then the ratio of $p_T$-integrated cross sections would
be affected but not the $p_T$ variation.}
{\em The $p_T$-dependence for such ratios is then a direct probe 
of the QGP physics.}

\section{Sensitivity of quarkonium suppression to the QGP model}

\indent\indent
As we have just shown, the $\Upsilon$ production rate is high enough at the LHC
for any QGP effects on the $\Upsilon$ family to be measurable.  Our discussion
in Sec.~2 suggested that at the high temperatures predicted for the minijet
plasma, $\Upsilon$ production may be suppressed at the LHC.  We would like to
use this suppression to determine the initial conditions of the QGP,
including $T_0$, $t_0$, and 
the temperature dependence of the QGP screening mass $\mu(T)$.  
A simple measurement of the overall suppression of the production
rates for the Upsilon family members will not be adequate.  We must
consider additional observables that could be affected by the presence
of the QGP. In this section, we will show that the $p_T$-dependence
of the suppression provides a remarkable amount of information.

If the QGP were of infinite temporal and spatial extent, resonance production
would be suppressed as long as $T>T_D$. But, there would be no significant
$p_T$-dependence.  In fact, however, both the lifetime and the
spatial extent of the QGP are finite.
Since a color singlet $Q \overline Q$ pair takes some time to form a bound 
state, the competition between the formation time, the time needed for the 
$Q \overline Q$ to
separate to the bound state radius\footnote{If the $Q \overline Q$ is produced
as a color octet, it is less likely to produce quarkonium than heavy-flavored
hadrons.}, and the characteristics of the plasma all 
affect the $p_T$-dependence of the $\Upsilon$ states.  
We will now discuss the sensitivity of the onium suppression
patterns as a function of $p_T$ to these ingredients.

The finite lifetime of the plasma sets an upper limit on the $p_T$ at which the
onium states can be suppressed.  In the plasma rest frame, the $b \overline b$
forms a bound state in the time $t_F = \gamma \tau_F$ 
where $\gamma = \sqrt{1 +
(p_T/M)^2}$ ($M$ is the onium mass) and the formation times, $\tau_F$,
for the bottonium and charmonium states were given in Table \ref{tablei}.   
When the plasma has cooled below $T_D$, the production of the
bound state is no longer suppressed.  The time at which this occurs can be
determined from the Bjorken model \cite{Bjorken} which assumes that the
evolution of the system is described by ideal hydrodynamics, cooling 
by an isentropic expansion so that
\begin{eqnarray}    
s_D t_D = s_0 t_0 \, \, . 
\label{entro}
\end{eqnarray}
Since the entropy density, $s$, is proportional to $T^3$, the time at which
the temperature drops below $T_D$ is
\begin{eqnarray} t_D = t_0 \left( \frac{T_0}{T_D} \right)^3 \, \, . 
\label{td}
\end{eqnarray} 
As long as $t_D/t_F > 1$, quarkonium formation will be
suppressed.  Thus the maximum $p_T$ at which suppression occurs is
\begin{eqnarray} 
p_{T, {\rm max}} = M \sqrt{(t_D/\tau_F)^2 - 1} \, .
\label{ptmax}
\end{eqnarray} 
The values for $t_D$ and $p_{T,{\rm max}}$ 
for all the onium states are given in Tables~\ref{tdpttablelhc}
and \ref{tdpttablerhic} for both forms of $\mu(T)$, eqs.~(\ref{mut1}) and
(\ref{mut2}), and three different
sets of initial conditions for LHC and RHIC energies.

\begin{table}[htbp]
{\small
\begin{center}
\begin{tabular}{|c|c|c|c|c|} \hline
\multicolumn{5}{|c|}{LHC} \\ \hline\hline
& \multicolumn{2}{c|}{SU($N$) only, eq.~(\ref{mut1})}  & 
\multicolumn{2}{c|}{gluons and quarks, eq.~(\ref{mut2})} \\ \hline
\multicolumn{5}{|c|}{minijet initial conditions} \\ \hline
& \multicolumn{2}{c|}{$T_0=1.14$ GeV, $t_0=0.1$ fm}  & 
\multicolumn{2}{c|}{$T_0=900$ MeV, $t_0=0.1$ fm} \\ \hline
 & $t_D$ (fm) & $p_{T, {\rm max}}$ (GeV) &  $t_D$ 
(fm) & $p_{T, {\rm max}}$ (GeV) \\ \hline
$\Upsilon$ & 2.49 & 29.5 & 0.105 & 0 \\ \hline
$\Upsilon'$ & 8.474 & 43.5 & 1.67 & 0 \\ \hline
$\chi_b$ & 8.474 & 30.7 & 1.46 & 0 \\ \hline
$\psi$  & 8.474 & 29.0 & 1.54 & 4.34 \\ \hline
$\psi'$ & 8.474 & 20.6 & 15.7 & 38.6 \\ \hline
$\chi_c$ & 8.474 & 14.4 & 17.5 & 30.5 \\ \hline
\multicolumn{5}{|c|}{Uncertainty relation initial conditions} \\ \hline
\multicolumn{5}{|c|}{$C=1$} \\ \hline
& \multicolumn{2}{c|}{$T_0=600$ MeV, $t_0=0.11$ fm}  & 
\multicolumn{2}{c|}{$T_0=785$ MeV, $t_0=0.084$ fm} \\ \hline
\multicolumn{5}{|c|}{$C=3$} \\ \hline
& \multicolumn{2}{c|}{$T_0=346$ MeV, $t_0=0.57$ fm}  & 
\multicolumn{2}{c|}{$T_0=453$ MeV, $t_0=0.436$ fm} \\ \hline
 & $t_D$ (fm) & $p_{T, {\rm max}}$ (GeV) &  $t_D$ 
(fm) & $p_{T, {\rm max}}$ (GeV) \\ \hline
$\Upsilon$ & 0.39 & 0 & 0.03 & 0 \\ \hline
$\Upsilon'$ & 1.3 & 0 & 0.455 & 0 \\ \hline
$\chi_b$ & 1.3 & 0 & 0.82 & 0 \\ \hline
$\psi$  & 1.3 & 3.27 & 0.42 & 0 \\ \hline
$\psi'$ & 1.3 & 0 & 4.28 & 9.9 \\ \hline
$\chi_c$ & 1.3 & 0 & 4.78 & 7.6 \\ \hline
\multicolumn{5}{|c|}{parton gas initial conditions} \\ \hline
& \multicolumn{2}{c|}{$T_0=820$ MeV, $t_0=0.5$ fm}  & 
\multicolumn{2}{c|}{$T_0=820$ MeV, $t_0=0.5$ fm} \\ \hline
 & $t_D$ (fm) & $p_{T, {\rm max}}$ (GeV) &  $t_D$ 
(fm) & $p_{T, {\rm max}}$ (GeV) \\ \hline
$\Upsilon$  & 4.6   & 56.53 & 0.19  & 0     \\ \hline
$\Upsilon'$ & 15.69 & 81.98 & 3.09  & 12.83 \\ \hline
$\chi_b$    & 15.69 & 58.9  & 5.53  & 18.58 \\ \hline
$\psi$      & 15.69 & 54.0  & 2.85  & 9.34  \\ \hline
$\psi'$     & 15.69 & 38.5  & 29.1  & 71.6  \\ \hline
$\chi_c$    & 15.69 & 27.2  & 32.5  & 56.76 \\ \hline
\end{tabular}
\end{center}
}
\caption[]{LHC values of $t_D$, and $p_{T, {\rm max}}$ for the two
choices, eq.~(\ref{mut1}) and eq.~(\ref{mut2}),
of $\mu(T)$, and for three sets of initial conditions.} 
\label{tdpttablelhc}
\end{table}

\begin{table}[htbp]
{\small
\begin{center}
\begin{tabular}{|c|c|c|c|c|} \hline
\multicolumn{5}{|c|}{RHIC} \\ \hline\hline
& \multicolumn{2}{c|}{SU($N$) only, eq.~(\ref{mut1})}  & 
\multicolumn{2}{c|}{gluons and quarks, eq.~(\ref{mut2})} \\ \hline
\multicolumn{5}{|c|}{minijet initial conditions} \\ \hline
& \multicolumn{2}{c|}{$T_0=445$ MeV, $t_0=0.1$ fm}  & 
\multicolumn{2}{c|}{$T_0=360$ MeV, $t_0=0.1$ fm} \\ \hline
 & $t_D$ (fm) & $p_{T, {\rm max}}$ (GeV) &  $t_D$ 
(fm) & $p_{T, {\rm max}}$ (GeV) \\ \hline
$\Upsilon$ & 0.15 & 0 & 0.003 & 0 \\ \hline
$\Upsilon'$ & 0.5 & 0 & 0.05 & 0 \\ \hline
$\chi_b$ & 0.5 & 0 & 0.094 & 0 \\ \hline
$\psi$  & 0.5 & 0 & 0.048 & 0 \\ \hline
$\psi'$ & 0.5 & 0 & 0.493 & 0 \\ \hline
$\chi_c$ & 0.5 & 0 & 0.55 & 0 \\ \hline
\multicolumn{5}{|c|}{Uncertainty relation initial conditions} \\ \hline
\multicolumn{5}{|c|}{$C=1$} \\ \hline
& \multicolumn{2}{c|}{$T_0=400$ MeV, $t_0=0.164$ fm}  & 
\multicolumn{2}{c|}{$T_0=524$ MeV, $t_0=0.126$ fm} \\ \hline
\multicolumn{5}{|c|}{$C=3$} \\ \hline
& \multicolumn{2}{c|}{$T_0=231$ MeV, $t_0=0.85$ fm}  & 
\multicolumn{2}{c|}{$T_0=303$ MeV, $t_0=0.65$ fm} \\ \hline
 & $t_D$ (fm) & $p_{T, {\rm max}}$ (GeV) &  $t_D$ 
(fm) & $p_{T, {\rm max}}$ (GeV) \\ \hline
$\Upsilon$ & 0.17 & 0 & 0.013 & 0 \\ \hline
$\Upsilon'$ & 0.59 & 0 & 0.2 & 0 \\ \hline
$\chi_b$ & 0.59 & 0 & 0.364 & 0 \\ \hline
$\psi$  & 0.59 & 0 & 0.188 & 0 \\ \hline
$\psi'$ & 0.59 & 0 & 1.92 & 2.94 \\ \hline
$\chi_c$ & 0.59 & 0 & 2.14 & 1.31 \\ \hline
\multicolumn{5}{|c|}{parton gas initial conditions} \\ \hline
& \multicolumn{2}{c|}{$T_0=550$ MeV, $t_0=0.7$ fm}  & 
\multicolumn{2}{c|}{$T_0=550$ MeV, $t_0=0.7$ fm} \\ \hline
 & $t_D$ (fm) & $p_{T, {\rm max}}$ (GeV) &  $t_D$ 
(fm) & $p_{T, {\rm max}}$ (GeV) \\ \hline
$\Upsilon$  & 1.95   & 22.3 & 0.082  & 0     \\ \hline
$\Upsilon'$ & 6.63 & 33.4 & 1.3  & 0 \\ \hline
$\chi_b$    & 6.63 & 23.2 & 2.33  & 0 \\ \hline
$\psi$      & 6.63 & 22.6 & 1.2  & 2.8  \\ \hline
$\psi'$     & 6.63 & 15.9 & 12.3  & 30.12  \\ \hline
$\chi_c$    & 6.63 & 11.1 & 13.7  & 23.75 \\ \hline
\end{tabular}
\end{center}
}
\caption[]{RHIC values of $t_D$, and $p_{T, {\rm max}}$ for the two
choices, eq.~(\ref{mut1}) and eq.~(\ref{mut2}),
of $\mu(T)$, and for three sets of initial conditions.} 
\label{tdpttablerhic}
\end{table}

It is
easy to check whether or not the bottonium states are suppressed given the
formation time and $T_0$ from the minijet initial conditions.  The SU($N$)
parameterization of eq.~(\ref{mut1}), 
with $T_c = 260$ MeV, results in $t_D/\tau_F > 1$ for all the members of the 
Upsilon family.
Thus, all the states will be suppressed up to rather large 
values of $p_T$, suggesting that any
suppression should be easily observable within the $p_T$ acceptance of CMS.
However, $T_0$ is reduced to 900 MeV when quark degrees of
freedom are added as in the 3-Flavor scenario of eq.~(\ref{mut2}); 
the result is that $t_D/\tau_F < 1$ for all the bottonium states and no
suppression occurs.  In other words, if the 3-Flavor
quark-gluon plasma is finite
and quickly equilibrated, none of the $\Upsilon$ states will be suppressed, and
the $p_T$-dependence should be unchanged from $pp$ production modulo 
any nuclear effects on $\Upsilon$ production.  

In contrast, the charmonium states are suppressed in both 
the SU($N$) and the 3-Flavor $\mu(T)$ scenarios.  The
SU$(N)$ case gives values of $p_{T, {\rm max}}$ similar to or smaller than 
those of the bottonium family because of the smaller charmonium masses.
This should be contrasted with the 3-Flavor QGP for which the $J/\psi$
would be suppressed over a very narrow range of $p_T$ relative to the $\chi_c$
and $\psi'$ since the value of $T_D$ for the $J/\psi$ is nearly twice that of
the other charmonium states (see Table \ref{tdtable} for $T_D$).  
If the $p_T$-dependence
of the nuclear effects on the charmonium states is weak, then the ratio of 
higher charmonium states to the $J/\psi$ 
as a function of $p_T$ would still be useful.

We discuss the other initial conditions shown in Tables \ref{tdpttablelhc} and
\ref{tdpttablerhic} and their consequences later.

In a finite system, such as those produced in heavy-ion collisions, the
suppression also depends on the size of the system.  We present a schematic
calculation of the $p_T$-dependence of the suppression following Ref.\
\cite{ChuM}.  Similar analyses of QGP production have also been done
\cite{KS,BO,KP}.  We can assume that the initial entropy profile is dependent
on the radius of the QGP, 
\begin{eqnarray} 
s_0(r) = s_0 \left( 1 - 
\left(\frac{r}{R} \right)^2\right)^\beta \, \, , 
\label{s0r}
\end{eqnarray} 
so that
\begin{eqnarray} 
t_D(r) = t_D(0) \left( 1 - 
\left(\frac{r}{R} \right)^2\right)^\beta \, \, , 
\label{tdr}
\end{eqnarray} 
where $t_D(0)$ is the value of $t_D$ calculated in eq.~(\ref{td}) for
resonances produced in the center of the system and $R$ is the transverse 
radius of the plasma, assumed to be
no larger than the radius of the nucleus, $R \simeq 1.2A^{1/3}$.  
At any time $t$ during the evolution of the QGP, the spatial
boundary of the screening region is located at
the transverse radius $r_S$ given by $t_D(r_S) = t$, 
\begin{eqnarray} r_S = R 
\left( 1 - \left(\frac{t}{t_D(0)} \right)^{1/\beta}\right)^{1/2} \, \, .
\label{rs}
\end{eqnarray}  

In general, if the $Q\anti Q$ pair is created with transverse
position $x^\mu = (0,{\bf r},0)$ and momentum $p^\mu = (\sqrt{M^2 +
p_T^2},{\bf p_T},0)$, at the formation time $\tau_F$ the
pair will form a bound state at $x^\mu = (\tau_F\sqrt{1+(p_T/M)^2}, 
{\bf r} + \tau_F {\bf p_T}/M,0)$.
If $| {\bf r} + \tau_F {\bf p_T}/M | \geq r_S$, the $Q\anti Q$
pair escapes the QGP before $\tau_F$ and will
successfully form bound states.  This inequality
will be satisfied (and the pairs will escape)
for a range of angles between $\bf r$ and $\bf p_T$
--- $0\leq\theta \leq \theta_{\rm max}(r,p_T)$ --- 
provided $p_T$ is such that $(r+\tau_F p_T/M)>r_S$. 
Here, $\theta_{\rm max}$ 
also depends on $\tau_F$, $M$, and $r_S$ as detailed below.
The $p_T$-dependent probability that the $Q\anti Q$ pair survives
and is able to form an onium state,
$S(p_T)$, is the ratio of the number of
bound states produced by those pairs that escape the plasma 
to the maximum possible number of 
onium bound states that would be formed at the given $p_T$ in the absence
of the QGP:
\begin{eqnarray}
S(p_T) = \frac{\int_0^R dr r \rho(r) \theta_(r,p_T)}
{\pi\int_0^R dr r \rho(r)} \, \,, 
\label{spt}
\end{eqnarray} 
where we parameterize $\rho(r) = (1 - (r/R)^2)^\alpha$ and 
\begin{eqnarray}
\theta(r,p_T) = \left\{ \begin{array}{ll} \pi & \mbox{$z\leq -1$} \\
\cos^{-1} z & \mbox{$|z| < 1$} \\ 0 & \mbox{$z \geq 1$} \end{array}
\right. \, \, , 
\label{thetar}
\end{eqnarray} 
where 
\begin{equation}
z = {r_S^2 - r^2 - \left({\tau_F p_T\over M}\right)^2 \over
2\tau_F r {p_T\over M}}\,.
\end{equation}
When $z \geq 1$ the pair cannot escape, whereas when $z \leq -1$ the
pair always escapes.  

\begin{figure}[htb]
\let\normalsize=\captsize   
\begin{center}
\centerline{\psfig{file=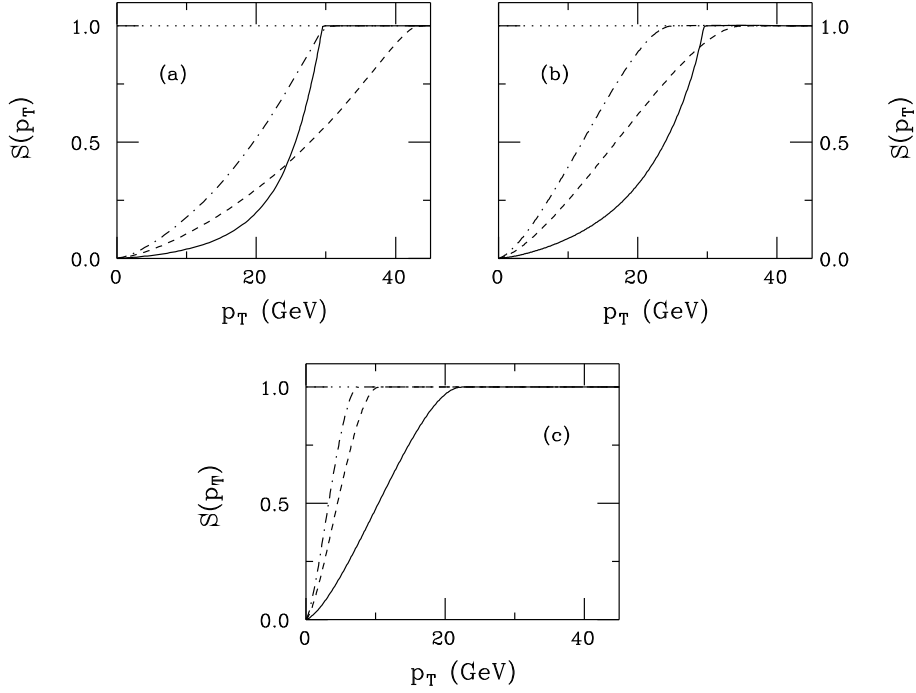,width=12.2cm}}
\begin{minipage}{12.5cm}       
\caption{For the SU($N$) $\mu(T)$ of eq.~(\protect\ref{mut1})
and the associated minijet initial conditions of 
$T_0=1.14$ GeV, $t_0=0.1$ fm, 
the suppression factor $S(p_T)$, calculated according to
eq.~(\protect\ref{spt}), is shown  for $\Upsilon$ (solid), $\Upsilon'$
(dashed), and $\chi_b$ (dot-dashed) production in three cases:
(a) $R=R_{\rm Pb}$; (b) $R=R_{\rm S}$; and (c) $R=1$ fm.}
\label{figuresptsun}
\end{minipage}
\end{center}
\end{figure}

The results for the survival probability $S(p_T)$,
with $\alpha = 1/2$ and $\beta = 1/4$, 
are shown in Figs.~\ref{figuresptsun}(a), \ref{figuresptsun}(b), and
\ref{figuresptsun}(c) 
for the $T_D$ values implied by eq.~(\ref{mut1})
(see Table~\ref{tdtable}) and for a decreasing sequence
of radii: (a) $R=R_{\rm Pb}=7.1$ fm, (b) $R=R_{\rm S}=3.8$ fm and (c) $R=1$ fm,
respectively. In all three cases, we observe that $S(p_T)\sim 0$,
corresponding to essentially total suppression, at small $p_T$ with our
assumption of an equilibrated system.
For $R=R_{\rm Pb}$ the
system is large enough for the finite lifetime to be more important than its
finite size.  Therefore $S(p_T) = 1$ for $p_T \geq p_{T, {\rm max}}$
while states with  $p_T < p_{T, {\rm max}}$ are suppressed.  If the QGP
is of smaller spatial extent, the survival probability can become
unity for $p_T < p_{T, {\rm max}}$ since a smaller
$p_T$ is sufficient for the pair to escape the system.  This
is illustrated by the results for $R=R_{\rm S}$ and $R=1$ fm
displayed in (b) and (c) of Fig.~\ref{figuresptsun}.
As $R$ decreases, the largest $p_T$ value for which
suppression occurs decreases more rapidly for the $\Upsilon'$ and $\chi_b$
states than for the $\Upsilon$ because of their larger breakup time, 
$t_D$; see Table~\ref{tdpttablelhc}.  In fact, when $R = 1$ fm,
$S(p_T) = 1$ for the $\Upsilon'$ at $p_T \approx 10$ GeV while the 
$\Upsilon$ remains suppressed up to $p_T \approx 20$ GeV, as shown in 
Fig.~\ref{figuresptsun}(c).

\begin{figure}[htb]
\let\normalsize=\captsize   
\begin{center}
\centerline{\psfig{file=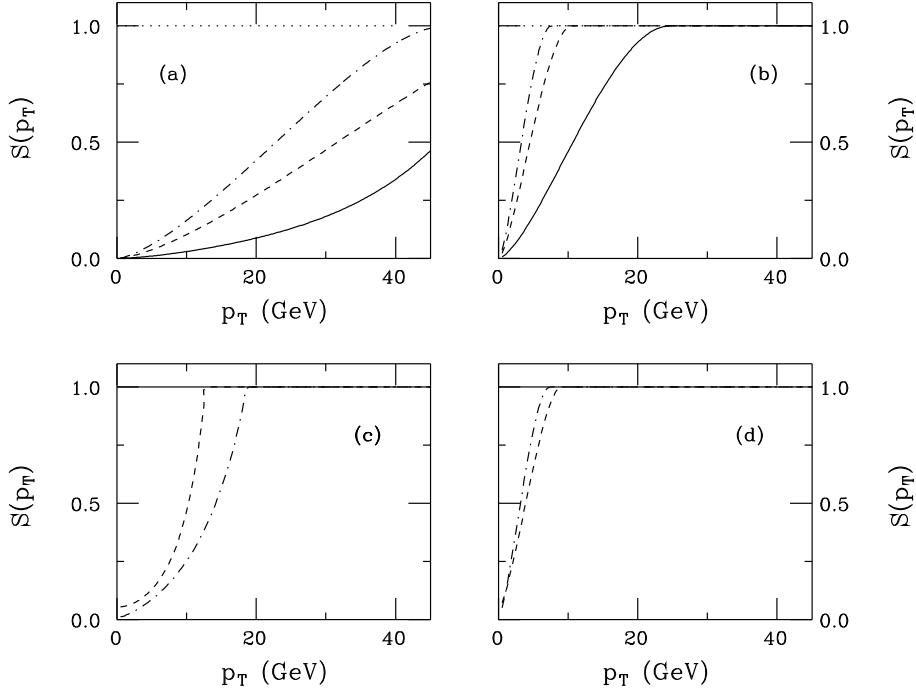,width=12.2cm}}
\begin{minipage}{12.5cm}       
\caption{For parton gas initial conditions of $T_0=820$ MeV, $t_0=0.5$ fm, 
the suppression factor $S(p_T)$, calculated according to
eq.~(\protect\ref{spt}), is shown  for $\Upsilon$ (solid), $\Upsilon'$
(dashed), and $\chi_b$ (dot-dashed) production in four cases:
(a) $R=R_{\rm Pb}$ and $\mu(T)$ of eq.~(\protect\ref{mut1});
(b) $R=1$ fm and $\mu(T)$ of eq.~(\protect\ref{mut1});
(c) $R=R_{\rm Pb}$ and $\mu(T)$ of eq.~(\protect\ref{mut2});
(d) $R=1$ fm and $\mu(T)$ of eq.~(\protect\ref{mut2}).}
\label{figuresptgas}
\end{minipage}
\end{center}
\end{figure}

\indent\indent We should compare these results to other possible QGP scenarios.
\begin{itemize}
\item
First, there is the possibility that we have overestimated
the amount of screening that occurs for a given level of minijet
production.  As we have noted, see Table~\ref{tdpttablelhc},
the screening times, $t_D$, are generally shorter for the perturbative 
estimate of eq.~(\ref{mut2}) [even
if quarks are not included by setting $n_f=0$] than for the SU($N$)
gauge parameterization given in eq.~(\ref{mut1})\footnote{The $\psi'$ and
$\chi_c$ are an exception because the lower value of $T_c$ found when quarks
are included reduced the breakup temperature, resulting in a longer screening
time.}.  The result is that there is no suppression of the
$\Upsilon$ states for the 3-Flavor QGP in equilibrium if minijet initial
conditions apply.
\item
It is also possible that the minijet density is different than we
assumed in our calculations. For example,
if the appropriate minijet momentum scale, $p_0$, were larger
(smaller) than the value of $p_0=2$ GeV assumed above, $t_0$ 
and, consequently, the minijet density, $T_0$ and, from eq.~(\ref{td}),
$t_D$ would all be smaller (larger).
A lower minijet density could produce
a different suppression pattern for the $\Upsilon$ family; \eg
a choice of $p_0$ is possible such that 
$\Upsilon$ production is not suppressed, while 
$\Upsilon'$ and $\chi_b$ production are.
\item
We note that typical early estimates of
the initial conditions without minijet production,
$t_0 = 1$ fm and $T_0 =250-300$ MeV at the LHC (see {\it e.g.}\ \cite{Satz91}),
produce no bottonium suppression regardless of the form of $\mu(T)$.
\item
An alternative set of initial conditions is obtained using the
uncertainty principle \cite{JLD,VJMR}.  In this case, 
$t_0 \langle E \rangle \sim C \hbar c$, with
$\langle E \rangle \sim 3T_0$ for a thermal average.  In previous calculations,
$C = 1$ and 3 were chosen.  Those studies, using $T_c = 200$ MeV
and assuming a 3-Flavor QGP, obtained $T_0 = 0.46$ GeV and $t_0 = 0.42$ fm
for $C=3$ and $T_0 = 0.81$ GeV and $t_0 = 0.08$ fm for $C=1$ \cite{VJMR}.
In Tables \ref{tdpttablelhc} and \ref{tdpttablerhic} we update these results
for the pure gluon SU($N$) and 3-Flavor QGP cases,
using the appropriate $T_c$ for each. As can be seen
from the formulae of Ref.~\cite{VJMR}, the 3-Flavor scenario
has a higher $T_0$ than the SU$(N)$ case given our assumption that 
the final multiplicity is the same regardless of the initial conditions.
We note that either choice of $C$ produces the same results because
$T_0 \propto \sqrt{3/C}$ (as shown in \cite{VJMR}) and $t_0 T_0 \propto C/3$,
implying that the $C$
dependence cancels in the expression for $t_D$, eq.~(\ref{td}).  
Neither scenario leads to
any bottonium suppression, and suppression of the charmonium states
is only possible for rather low $p_T$ values.
\item
The final initial condition scenario that we discuss is based on a
parton gas model as adopted for $J/\psi$ suppression at RHIC and LHC
energies in Ref.~\cite{XKSW}. In this model one finds that
the kinetic equilibration time is longer than for any of the cases
discussed so far, $t_0 \sim 0.5-0.7$ fm.  This time is reached
when the momentum distributions are locally isotropic due to elastic
scatterings and the expansion of the system.  Chemical equilibrium is generally
not assumed but the system moves toward equilibrium as a function of time.  In
this picture, the cooling of the plasma is more rapid than the simple Bjorken
scaling picture \cite{Bjorken} we have adopted, producing incomplete
suppression at low $p_T$.  To make a schematic comparison with our assumption
of complete equilibrium, we give in Tables
\ref{tdpttablelhc} and \ref{tdpttablerhic}
the values of $t_D$ and $p_{T, {\rm max}}$
for the onium states computed using the same $T_0$ and $t_0$ 
as employed in their calculation, derived from the 
HIJING Monte Carlo code \cite{HIJING}.  
(Note that we use the same initial values independent of 
the form assumed for $\mu(T)$.)
Because the equilibration time of the parton gas is longer than
that obtained from the minijet initial conditions, the time the system spends
above the breakup temperature is also longer, leading to suppression for both
the SU($N$) and the 3-Flavor
forms of $\mu(T)$, even though $T_0$ is lower.  These results are
illustrated in Fig.~\ref{figuresptgas} for the two $\mu(T)$ parameterizations.
Figs.~\ref{figuresptgas}(a) and \ref{figuresptgas}(b) illustrate 
the survival probability $S(p_T)$ for the SU($N$) plasma
assuming $R = R_{\rm Pb}$ and 1 fm, respectively.  In the SU($N$) case, 
$p_{T, \rm max}$ is sufficiently large 
that the lead nucleus is already small enough for the $p_T$ range of the
suppression to be less than the calculated $p_{T, {\rm max}}$.
The 3-Flavor QGP results are shown in 
Figs.~\ref{figuresptgas}(c) and \ref{figuresptgas}(d) 
for $R = R_{\rm Pb}$ and 1 fm, respectively.  Here the temperature is not
high enough for $\Upsilon$ suppression but the $\Upsilon'$ and $\chi_b$ are
suppressed, albeit over a much smaller $p_T$ range than in the SU($N$) case.
\end{itemize}

\begin{figure}[htb]
\let\normalsize=\captsize   
\begin{center}
\centerline{\psfig{file=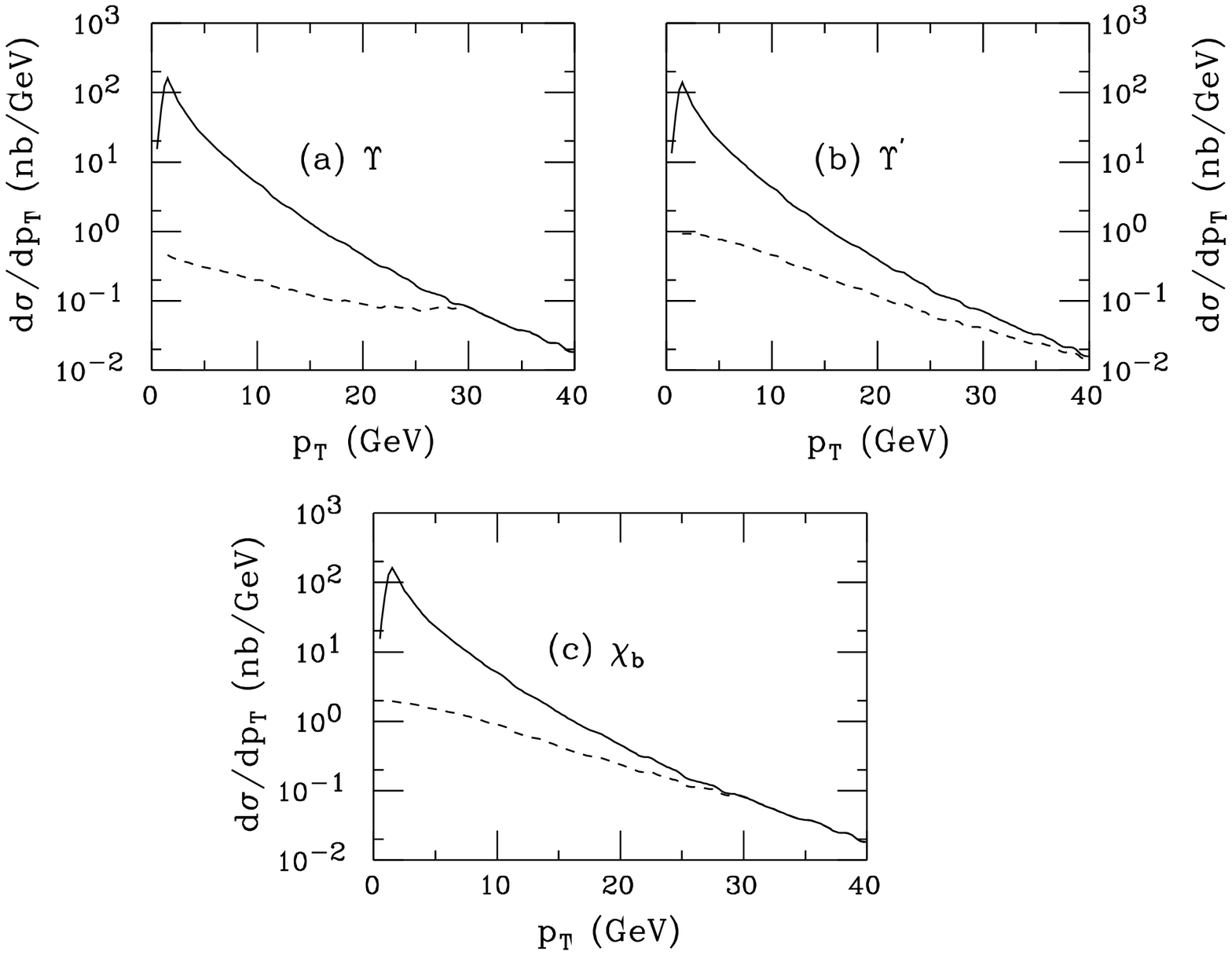,width=12.2cm}}
\begin{minipage}{12.5cm}       
\caption[]{
The $p_T$ distributions for (a) $\Upsilon$, (b) $\Upsilon'$,
and (c) $\chi_b$ production are given for $pp$ interactions (solid) and plasma
production (dashed).  The plasma results are the $pp$ distributions
multiplied by $S(p_T)$ as computed for $R=R_{\rm Pb}$,
$t_0=0.1$ fm and $T_0=1.14$ GeV and assuming the SU($N$) form
of $\mu(T)$ as given in eq.~(\ref{mut1}); $S(p_T)$ for this case was plotted
in Fig.~\ref{figuresptsun}(a). We have taken $\rts=5.5\tev$.
The differential cross
sections (before including $S(p_T)$) are normalized to the integrated
values tabulated in Table~\ref{tablerates}.}
\label{figuredsigdpt}
\end{minipage}
\end{center}
\end{figure}

A given prediction for the survival probability $S(p_T)$ can be used to 
calculate the $p_T$ distributions of the $\Upsilon$
states. Since $S(p_T)$ depends upon the $T_D$ and $\tau_F$ 
of each Upsilon resonance, each state will have a unique $p_T$
dependence, despite the fact
that the $p_T$ distributions determined from the color evaporation
model are the same for directly produced $\Upsilon$, $\Upsilon'$ and $\chi_b$
before inclusion of QGP effects.
As we have noted, the precise $S(p_T)$ factors for the different
$\Upsilon$ resonances also depend on 
the choice of $\mu(T)$, $t_0$, $T_0$, and $R$.
The resulting $p_T$ spectra are shown in Fig.~\ref{figuredsigdpt} in the
case of $S(p_T)$ taken from Fig.~\ref{figuresptsun}(a), assuming the SU$(N)$
form of $\mu(T)$, with
$R=R_{\rm Pb}$, $t_0=0.1$ fm and $T_0=1.14$ GeV.
The upper curves are the $p_T$ distributions in
$pp$ collisions at $\sqrt{s} = 5.5$ TeV while the lower ones are the $pp$
curves multiplied by $S(p_T)$ to indicate the $p_T$ distribution 
expected in the presence of the quark-gluon
plasma.  Note that plasma production decreases the slope of the $p_T$
distribution.  

The raw $p_T$ distributions before including the survival probabilities
were obtained using the color evaporation model predictions calculated
with the next-to-leading order $Q \overline Q$ code of Ref.\ \cite{MNR}.
It is well-known
that at low $p_T$ it is necessary to resum the next-to-leading
$\ln(M^2/p_T^2)$ powers for each power of $\alpha_s$ to get
an exponential factor.
This has been extensively studied for $W$ and $Z$
production \cite{css,dws,aegm,arnoldkauf,ladinskyuan,reno}. 
The procedure involves computing a form factor in impact parameter ($b$)
space, and then Fourier transforming back to $p_T$ space.
For $\lamqcd\ll 1/b<M$, the form factor can be computed perturbatively;
for $b\gsim 1/\lamqcd$ non-perturbative modeling is necessary.
In the latter range, the conventional procedure is to 
introduce a cutoff in $b$
and a non-perturbative factor in the $b$-space form factor. 
It is found \cite{reno} that the non-perturbative component dominates
the $p_T$-dependence modifications for $p_T\lsim 6\gev$, implying
substantial uncertainties for $p_T$-dependence in this domain.
However, Ref.~\cite{reno} also demonstrates that
the ratio $[d\sigma(W)/dp_T]/[d\sigma(Z)/dp_T]$ is very insensitive
to the non-perturbative uncertainties due to the fact
that $\mw$ is not very different from $\mz$.  Further, the ratio
deviates only slightly from the perturbatively computed value.
Applied to the closely-spaced resonances of the Upsilon family,
the formalism of Ref.~\cite{reno} implies that the color evaporation
model predictions for the 
ratios of the production rates for different $\Upsilon$ states 
will also not be significantly modified by the non-perturbative physics.  
The ability to reliably compute a ratio such as $\Upsilon'/\Upsilon$
as a function of $p_T$ down to relatively low
$p_T$ values is critical to the program described in the next section.

\section{A program for determining the nature of the QGP}

\indent\indent
We have seen that the uncertainties in the initial conditions
as reflected in $t_0$ and $T_0$, the uncertainty in the QGP radius $R$,
and the ambiguity in the temperature dependence
of the screening mass, $\mu(T)$, all imply that there are many
possible results for the relative suppression of the
$\Upsilon$ states. The challenge will be to extract the correct 
plasma picture from the data.  
We describe below the inputs, procedures and detector features
required in order that the production
and suppression of the $\Upsilon$ states can be used
as a tool to determine the true state of the plasma and the proper initial
conditions. Data from the LHC and RHIC will prove complementary;
it is likely that Upsilon family production rates
must be measured at both machines in order to reach reliable conclusions.
Further, measurements of the charmonium states at both LHC and RHIC 
will provide valuable information.
We organize the following discussion into a series
of (overlapping) tasks and requirements necessary to extract information about
the QGP.

\medskip
\noindent{\bf Checking the color evaporation model and establishing
a baseline.}
\smallskip

As a first step, the production rates and $p_T$
spectra for the $\Upsilon$ states must be measured
in $pp$ collisions at the standard LHC nucleon-nucleon
collision energy of $5.5$ TeV and, by lowering the energy of the LHC still
further, at $1.8$ TeV.\footnote{The lower energy measurement should agree
with that obtained at the Tevatron \cite{Eggert2}
to the extent that valence
quark distribution function contributions do not contribute significantly
in $p\anti p$ collisions, but a direct $pp$ measurement
will remove the necessity of making this assumption in
order to carry out the following program.}
This will allow us to check the predictions\footnote{Assuming that
gluon fusion is dominant at both LHC and Tevatron energies.} 
of the color evaporation model. In particular, we must confirm
that the the $\Upsilon'/\Upsilon$ and $\Upsilon''/\Upsilon$ ratios
are {\it independent} of energy and $p_T$ and that the (common) 
$p_T$-dependence of the individual cross sections agrees with the color 
evaporation prediction. These measurements will allow us
to confirm our perturbative modeling and understanding
of $\Upsilon$ production in the clean $pp$ environment, thus establishing a 
baseline for comparison with results obtained for nuclear collisions.

\medskip
\noindent{\bf Determining the effects of shadowing and nucleon absorption 
using RHIC data.}
\smallskip

Next, we must determine if the $\Upsilon'/\Upsilon$ ratio is independent
of purely nuclear matter effects (\eg shadowing and absorption) 
by measuring the ratio in $pA$ collisions.  Since these studies are not
possible at the LHC, this must be established at RHIC.  
The heaviest nuclear beam at RHIC will be gold, but the shadowing should be
quite similar in gold and lead.  Since RHIC is a dedicated heavy-ion facility,
the PHENIX experiment at RHIC is expected
to obtain several thousand $\Upsilon$ events per year along with
a similar number of $\Upsilon'$'s \cite{PHCDR}.  
This should be sufficient to determine: a) the $p_T$ dependence of the
$\Upsilon'/\Upsilon$ ratio in $pA$ interactions where
nuclear effects are present but no QGP is expected;
and b) the magnitude of the $\Upsilon'/\Upsilon$ ratio.
If the ratio is the same (smaller) than in $pp$ collisions,
then shadowing (absorption) is probably the dominant purely nuclear effect.
The dependence of the onium ratios on shadowing/absorption
may also be established by studying $\psi'/\psi$ in $pA$ collisions: 
lower $x$ partons are probed and the rates are higher.

If shadowing is dominant over absorption,
then measurements of the actual magnitude of the gluon shadowing will
be important if we are to use the absolute
$p_T$-dependence of the onium cross sections instead of their ratios
to analyze the QGP. It has been
suggested that $pA$ measurements of the dilepton continuum ($M \geq 1$ GeV)
at RHIC will provide a measure of gluon shadowing since dileptons
from charm production and decay (mainly from gluon fusion)
will be the largest contribution in the mass region below the $J/\psi$
\cite{LG2}.  Such charm studies are particularly useful because 
they probe gluon shadowing at the low $x$ values that will
be important for Upsilon production at the LHC. Measurements at RHIC
for dilepton masses near the
$\Upsilon$ mass probe parton $x$ values of $x \sim 0.1$ (at $y \sim 0$),
\ie substantially above those appropriate at the LHC and also
in a range where little shadowing is expected \cite{VGMR}.  
In order to probe small $x$ values for dilepton masses
near $M_{\Upsilon}$, it would be necessary to accumulate
good statistics for the $\Upsilon$'s at high rapidity, which can be
observed in the muon arms of the detector.
An $\Upsilon$ with $y \sim +2$ or $-2$ 
will have $x_2 \sim 0.35$, \ie in the EMC region, or $x_2 \sim 0.006$,
in the shadowing region, respectively. 
The latter would probe nuclear effects on $\Upsilon$ production 
for gluon $x$ values that are dominant for $\Upsilon$
production at the LHC around $y \sim 0$.

Perhaps the best way to determine the importance of nucleon absorption for the
onium states is to measure the relative rapidities of the onium states and the
initial nucleons since the baryon number in the central region of nuclear
collisions at RHIC and LHC is expected to be small.  
If the nucleons and quarkonium states are far apart in phase
space, the effects of nucleon absorption should be negligible.  A zero-degree
calorimeter or a forward hadron spectrometer, such as that of the BRAHMS
detector at RHIC \cite{BRAHMS} could clarify this matter.

\medskip
\noindent{\bf Determining the importance of comovers.}
\smallskip

The use of onium production data for determining the
nature of the QGP will be difficult if there are non-QGP mechanisms
(other than simple shadowing) affecting onium production 
that are comparable to the predicted QGP
effects. We have already discussed one such mechanism, namely nuclear
absorption.  However, we concluded that it was relatively
benign since it should have little $p_T$ dependence and will not be important
if the projectile and target fragmentation regions are well separated from the
quarkonium states in the central region. A second
mechanism of concern is the breakup/absorption of the onium states
through interactions with comoving secondary hadrons. Indeed,
the observed $J/\psi$ suppression \cite{NA38,NA50} at $\sqrt s=17\gev$
has been attributed to interactions with these comovers \cite{GV2}.  The
comovers should also be present at higher
energies.  Such hadronic scattering can affect both the $\psi$ states and the
$\Upsilon$ states.  For example, the comovers can easily break up the $J/\psi$
if the combined comover+$J/\psi$ center-of-mass energy is
above the $D\overline D$ threshold.  
Since the separation between the onium mass and the threshold is smaller 
for the higher onium states, the higher states will be more easily broken 
apart by comovers.  Thus, comovers 
could cause ratios such as $\Upsilon'/\Upsilon$ and $\psi'/\psi$
to be smaller than expected. The $\Upsilon$, being the most
tightly bound of the onium states, should be least affected.
Note that this is precisely the same as expected
from nucleon absorption, which should be largest for states with large radii
and small mass thresholds.
As discussed earlier, if the comovers are mesons, the onium-comover absorption
cross section should be $\sim 2/3$ the onium-nucleon absorption cross section.

Although the interpretation of $J/\psi$ suppression at $\sqrt s=17\gev$
as being due to comover scattering implies high hadronic densities in 
the present experiments, a hadronic interpretation of quarkonium dissociation
would be much less plausible at LHC energies, where, for all cases
considered in Table~\ref{tdpttablelhc}, $T_0$ is significantly above $T_c$. 
The reasoning
is as follows. For such high initial temperatures, it is almost certain
that a QGP will form, delaying the creation of hadrons until the $Q\anti Q$
pair or onium state has passed through much of the material.
Thus, there would be little opportunity for interactions with comovers until
the late stages of the collision.
The $T_0$ values for RHIC (Table~\ref{tdpttablerhic}) are also generally
higher than $T_c$ and significant breakup due to interactions
with comovers again seems unlikely if a QGP is formed.

While the above argument suggests that comover absorption
will not significantly
impact onium production, it will be important to try to determine
the importance of comovers directly from experiment. We outline several
possible approaches.

\begin{itemize}
\item
In $pA$ collisions a QGP is not formed. Thus,
as noted earlier, in $pA$ collisions the dilepton spectrum in the $M>1\gev$ 
region, being primarily due to $c\anti c$ production and decay,
should provide a reliable probe of shadowing.  Comovers will have only
a small influence on open charm mesons because the scattering is elastic
--- a
$\pi D$ interaction will change the momentum distribution of the $D$ but will
not take the charm out of the system. Nucleon-$D$ scattering is also expected
to be elastic. On the other hand, a $J/\psi$-comover
interaction is likely to break up the $J/\psi$ into $D \overline D$ pairs.
By comparing the `shadowing' on and off the resonance, we can
obtain a firmer handle on the magnitude of comover absorption of the $J/\psi$.
\item
In nucleus-nucleus collisions, the density of comovers which can absorb the
onium state
would be larger still, with the largest density in central collisions.
Since these densities are related to the final multiplicity and thus to the
transverse energy of the collision, perhaps one of the best ways to check the
importance of comover breakup and absorption is to trigger on
peripheral collisions (for which a QGP is not formed).
One would compare predictions for the $\psi$ and $\Upsilon$
states based on shadowing in peripheral collisions
with actual measurements to see if there
is extra suppression due to comover breakup or absorption.   Such collisions
could provide very valuable guidance on whether comover
absorption is important.  In addition,
one could compare (as above) the dilepton spectrum on and off the $\psi$
resonance.
\end{itemize}

Comover absorption will also not occur if the onium $p_T$ is large enough.
Calculations of comover suppression as a function of $p_T$
appear in Refs.~\cite{FLP,GGJ,VPKH}.  The range of $p_T$ over which
comover suppression occurs should not depend strongly on energy.  One can make
a crude estimate of the $p_T$ range of comover interactions as in
eq.~(\ref{ptmax}).  In this case,
\begin{equation}
p_{T, {\rm max}}^{\rm co} = M \sqrt{(t_{\rm co}/\tau_{\rm co})^2 - 1}
\, \, ,
\end{equation}
where $t_{\rm co}$ depends on the size of the system, $t_{\rm co} \sim R/v_{\rm
rel}$, ($v_{\rm rel}$ is the relative velocity of the onium and the comovers,
0.6 for $\psi$-comover interactions at the CERN SPS) 
and $\tau_{\rm co} \sim 2$ fm is the 
comover formation time in a nucleus-nucleus collision.  Since we assume
comovers are final-state hadrons, there is no $T_0$ dependence and since the
$p_T$ range depends on the comover formation time rather than the onium
formation time, $p_{T, {\rm max}}^{\rm co}$ changes by 20\% or less for each
onium family.  When $R=R_{\rm Pb}$,  $p_{T, {\rm max}}^{\rm co} = 5.8M$.  Note
that if a QGP is formed, $\tau_{\rm co}$ could be even longer since the
comovers would not form until after the QGP hadronizes.
At the very least, the above comparisons should allow us to determine
the maximum $p_T$ up to which comover
absorption is important.

The influence of comover absorption is clearly an important issue.  
In the following discussion, we assume that all nuclear and
comovers effects on the onium production cross sections can
be unfolded or proven to be small.

\medskip
\noindent{\bf Using RHIC data to establish the general nature of
the QGP.}
\smallskip

At RHIC ratios such as
$\Upsilon'/\Upsilon$ and $\psi'/\psi$ will be measured
in nucleus-nucleus collisions.  For minijet initial conditions
or initial conditions based on the uncertainty relations,
the short initial time scales result in the prediction that
there should be no significant suppression at RHIC for any of the onium
states, regardless of the form of $\mu(T)$.
However, in the parton gas model the long $t_0$ implies that suppression of
the $\Upsilon$ states is possible for the SU($N$) (but not
the 3-Flavor) form of $\mu(T)$.
Further, the $\psi$ states are predicted to be
suppressed for both the SU($N$) and 3-Flavor forms of $\mu(T)$.  
Thus, the suppression pattern of the $\psi$ and
$\Upsilon$ states will help distinguish the $\mu(T)$ dependence if significant
suppression is observed.  Specifically:
\begin{itemize}
\item
If all the ratios are independent of $p_T$, then
either no plasma has been produced or the initial conditions of the plasma
are unfavorable for suppression to occur.
\item
If the $\psi'/\psi$ ratio depends on $p_T$, but the $\Upsilon'/\Upsilon$
ratio does not, the most consistent picture is that
a QGP has formed with initial conditions like those
of the parton gas model and $\mu(T)$ is of the 3-Flavor form given
in eq.~(\ref{mut2}).  
\item
If both ratios are $p_T$-dependent, then parton-gas-like
initial conditions are required and the screening mass is
more likely to be of the SU$(N)$ form as in eq.~(\ref{mut1}).
\end{itemize}

\medskip
\noindent{\bf Analyzing the QGP at the LHC.}
\smallskip

One advantage of the LHC over RHIC is that onium suppression
can occur for a much broader range of possible initial conditions.
In particular, at the LHC the minijet initial conditions produce
an initial temperature $T_0$ that is sufficiently large 
that the fast equilibration predicted (\ie small $t_0$) 
does not prevent suppression. More generally,
the higher $T_0$ possible at the LHC will be crucial since fast
equilibration is perhaps more plausible.

The question we now address is
how to best determine $t_0$ and other features of
the QGP. A high statistics study of the
$\Upsilon'/\Upsilon$ ratio as a function of $p_T$ may prove conclusive. 
If significant $p_T$-dependence of $\Upsilon'/\Upsilon$ is
found for $p_T\gsim 10\gev$ in Pb+Pb collisions
at $5.5\tev$, then it will be virtually certain that a quark
gluon plasma was formed. (As argued at the end of the previous section, any
$p_T$-dependence of this type of ratio arising from 
low-$p_T$ perturbative or non-perturbative resummation effects
should be small, and effects from comovers/absorption/shadowing should 
be absent once $p_T$ is reasonably large.)
The precise behavior of the
$\Upsilon'/\Upsilon$ ratio can then be used to strongly constrain
the QGP model parameters.
In particular, the ratio will be very different if only
the $\Upsilon'$ is suppressed in comparison to the case
where both states are suppressed.

\begin{figure}[htb]
\let\normalsize=\captsize   
\begin{center}
\centerline{\psfig{file=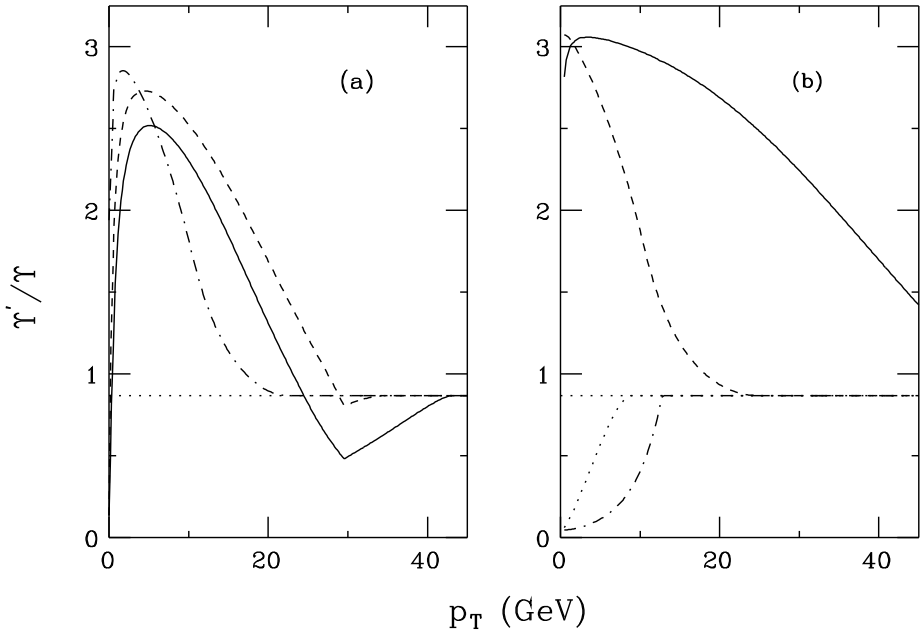,width=12.2cm}}
\begin{minipage}{12.5cm}       
\caption{
The direct or prompt $\Upsilon'/\Upsilon$ ratio 
as a function of $p_T$ is shown for
several choices of initial conditions at the LHC. In (a),
we show results for the minijet model with the SU($N$) form of $\mu(T)$,
eq.~(\protect\ref{mut1}),
(for which $T_0=1.14$ GeV and $t_0=0.1$ fm)
taking $R=R_{\rm Pb}$ (solid), $R=R_{\rm S}$ (dashed),
$R=1$ fm (dot-dashed). In (b), we consider the parton gas model with
$T_0=0.82$ GeV and $t_0=0.5$ fm, and
show results for the SU($N$) $\mu(T)$ with $R=R_{\rm Pb}$ (solid) and $R=1$
fm and for the 3-Flavor $\mu(T)$ from eq.~(\protect\ref{mut2}) 
with $R=R_{\rm Pb}$ (dot-dashed) and $R=1$ fm (dotted).}
\label{figureprompt}
\end{minipage}
\end{center}
\end{figure}

In Fig.~\ref{figureprompt}, we show the $\Upsilon'/\Upsilon$ ratio
for several sets of initial conditions. The 
`prompt' ratio displayed is that obtained if 
only directly produced $\Upsilon'$ and
$\Upsilon$ states are included. 
In other words, the ratio is computed
using $f^d_{\Upsilon'}$ and $f^d_{\Upsilon}$ from eq.~(\ref{fds}).
We will return shortly to the issue of whether or not
this prompt ratio can be experimentally measured. 

Fig.~\ref{figureprompt}(a) displays results obtained
when the minijet initial conditions are used with $t_0=0.1$ fm
and $\mu(T)$ is as given in eq.~(\ref{mut1}). 
Both the $\Upsilon$ and $\Upsilon'$
states are suppressed and the solid, dashed and dot-dashed
curves (for $R=R_{\rm Pb}$, $R_{\rm S}$ and $1$ fm,
respectively) show that the $p_T$-dependence of the 
ratio can distinguish different plasma radii.  
At low $p_T$ the $\Upsilon$ is actually more suppressed
than the $\Upsilon'$, leading to a significantly larger ratio than 
that expected in low energy $pp$ collisions (or found in
current $p\anti p$ measurements).
For a large $R$, the ratio eventually 
drops below the value expected in the absence of QGP effects
(\ie for large enough $p_T$).  
The point at which this shift occurs is at $p_T \approx
25$ GeV if $R = R_{\rm Pb}$ and at $\approx 29$ GeV for $R=R_{\rm S}$.  The
kink in these ratios at 30 GeV occurs at the point where the $\Upsilon$ is
no longer suppressed.  In a system 
with a smaller $R=1$ fm radius, the $\Upsilon$ is always
more suppressed than the $\Upsilon'$ so that the ratio returns to 
the $pp$ value at $p_T \approx 20$ GeV.  

Fig.~\ref{figureprompt}(b) shows the ratio computed using
the parton gas initial conditions.
When the SU$(N)$ $\mu(T)$ is assumed, the behavior is similar to the minijet 
case although when $R=R_{\rm Pb}$ the increase in $\Upsilon'/\Upsilon$ persists
to much larger $p_T$.  When $R=1$ fm, the result is almost indistinguishable
from the minijet case with the same radius.  The maximum of the ratio is
$\sim 10$\% larger at $p_T \sim 1-2$ GeV, but this may be difficult to
distinguish.  If the 3-Flavor [eq.~(\ref{mut2})] 
scenario is assumed for the screening mass,
then the $\Upsilon'$ is suppressed while the
$\Upsilon$ is not and the $p_T$-dependence of the ratio is quite different.
The dot-dashed and dotted curves of Fig.~\ref{figureprompt}(b)
show that until the effects of the plasma
disappear, the ratio is less than expected in the absence
of QGP formation. However,
it is more difficult to differentiate between a system the size of the
lead nucleus or a system with $R=1$ fm unless the measurement
is made with high statistics.  
We have already noted that if significant suppression is observed at RHIC
in the $\psi'/\psi$ ratio as a function of $p_T$ and $\Upsilon$ suppression is
observed at RHIC, the general behavior of $\mu(T)$ should be relatively clear
by the time the LHC measurement is made.

Finally, we recall that for the uncertainty relation initial conditions,
both $t_0$ and $T_0$ are small even at the LHC and no suppression
is expected for the Upsilon states.  To learn about the existence
and nature of a QGP via measurements of onium production will not be 
possible for such models.

\medskip
\noindent{\bf The importance of separating prompt from indirectly
produced onium states.}
\smallskip

\begin{figure}[htb]
\let\normalsize=\captsize   
\begin{center}
\centerline{\psfig{file=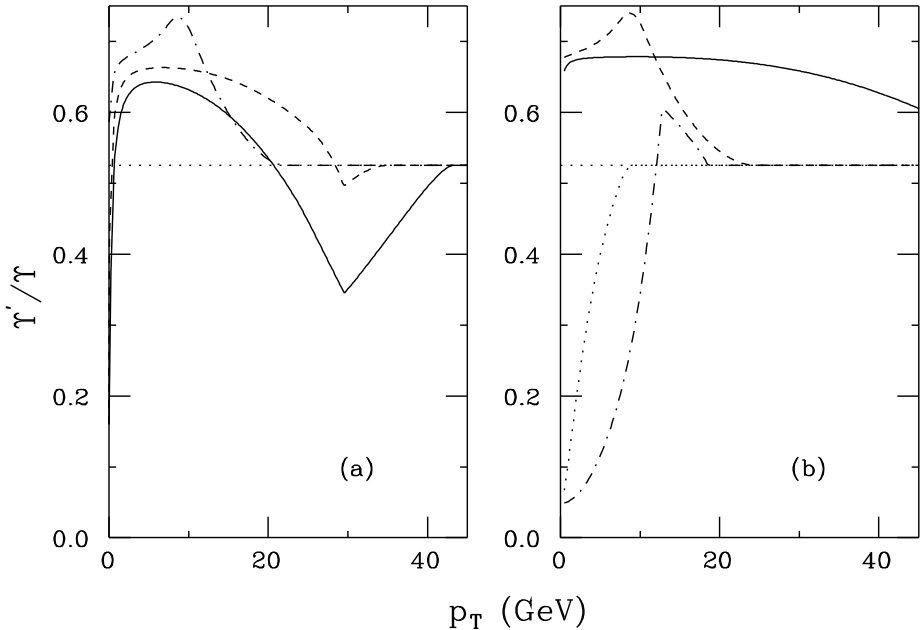,width=12.2cm}}
\begin{minipage}{12.5cm}       
\caption{
The $\Upsilon'/\Upsilon$ ratio including indirect decays
as a function of $p_T$ is shown for
several choices of initial conditions at the LHC. In (a),
we show results for the minijet model with the SU($N$) form of $\mu(T)$,
eq.~(\protect\ref{mut1}),
(for which $T_0=1.14$ GeV and $t_0=0.1$ fm)
taking $R=R_{\rm Pb}$ (solid), $R=R_{\rm S}$ (dashed),
$R=1$ fm (dot-dashed). In (b), we consider the parton gas model with
$T_0=0.82$ GeV and $t_0=0.5$ fm, and
show results for the SU($N$) $\mu(T)$ with $R=R_{\rm Pb}$ (solid) and $R=1$
fm and for the 3-Flavor $\mu(T)$ from eq.~(\protect\ref{mut2}) 
with $R=R_{\rm Pb}$ (dot-dashed) and $R=1$ fm (dotted).}
\label{figureindirect}
\end{minipage}
\end{center}
\end{figure}

An important issue is whether or not the `prompt' ratio discussed
above can be extracted experimentally. We saw earlier that a significant
fraction of $\Upsilon$ production arises from $\Upsilon'$ and $\chi_b$
($\chi_b$ denotes
the collective $\chi_{bi}(1P,2P)$, $i=0,1,2$ states) production
followed by $\Upsilon'\to\Upsilon X$ and
$\chi_b\to \gam\Upsilon$, respectively. Further, some of the $\Upsilon'$'s
come from $\chi_{bi}(2P)\to \gam \Upsilon'$.
We can only isolate the prompt ratio discussed above
if we can veto such decay sources of the $\Upsilon$ and $\Upsilon'$.
In the case of the $\chi_b$ decays, 
this will only be possible by detecting the associated
photon and reconstructing the $\chi_b$.  The ability of the LHC
detectors to do this will be limited at low $p_T$ for two reasons:
i) there is a large $\pi^0$ background from which it is
difficult to extract isolated $\gam$'s until the $p_T$
of the $\gam$ becomes substantial ($\gsim 10\gev$) and ii) the
photon energy resolution will be poor at low $p_T$.
In order to veto $\Upsilon'\to\Upsilon X$ decays we must be able
to detect the $X=\pi^+\pi^-,\pi^0\pi^0$ decay products.  This latter
may be impossible; however, this indirect decay is
only responsible for about 13\% of the $\Upsilon'$'s, so that
if the $\chi_b$ decays can be vetoed we will obtain a result much
like the prompt ratio plotted in Fig.~\ref{figureprompt}.  If the $\chi_b$
decays cannot be vetoed, then we must consider all sources
of the $\Upsilon'$ and $\Upsilon$, each of which will
be associated with a different suppression factor.
Numerically, it is sufficient to consider:
\begin{equation}
{\Upsilon'+\chi_b(2P)(\to \Upsilon')+\Upsilon''(\to\Upsilon')
\over
\Upsilon+\chi_b(1P,2P)(\to\Upsilon)+
\Upsilon'(\to\Upsilon)+\Upsilon''(\to\Upsilon)}\,.
\label{chain}
\end{equation}
In computing this `indirect' $\Upsilon'/\Upsilon$ ratio we have assumed
that the survival probability, $S(p_T)$, of the $\chi_b(2P)$ states is the same
as that for the $\chi_b(1P)$ states and that $S(p_T)$
for the $\Upsilon''$ is the same as for the $\Upsilon'$.

The indirect ratio, defined in eq.~(\ref{chain}) is plotted in
Fig.~\ref{figureindirect}. Clearly, the indirect ratio
is somewhat less sensitive to changes in $R$
and $\mu(T)$ than is the direct $\Upsilon'/\Upsilon$ ratio.
If the SU($N$) form of $\mu(T)$ is appropriate, the figure indicates that
a measurement at the 20\% level is needed to distinguish between the $pp$
value of the ratio and the QGP prediction including indirect decays
in both the minijet and parton gas initial condition scenarios.
(If there are substantial systematic errors
in the ratio, the detection of a deviation might be most difficult
for the parton gas SU($N$) case when $R=R_{\rm Pb}$,
since in this case the ratio varies slowly with $p_T$.)
If the slowly-growing 3-Flavor form of $\mu(T)$
is correct, the dot-dash and dotted
curves of Fig.~\ref{figureindirect}(b) show that the parton gas model
predicts significant and strongly $p_T$-dependent departures
from the $\Upsilon'/\Upsilon$ ratio expected in the absence
of a QGP. It would even be possible to determine the
size of the system using the indirect ratio --- as shown  
in Fig.~\ref{figureindirect}(b) the
$\chi_b$ contribution to the $\Upsilon$ causes the $\Upsilon'/\Upsilon$ ratio
to increase above the $pp$ level at $p_T \approx 12$ GeV for $R=R_{\rm Pb}$
but not for $R=1$ fm.  

Thus, all is not lost if only the indirect ratio can be measured,
but sensitivity is reduced and the accuracy with which the
$\Upsilon'/\Upsilon$ ratio can be measured becomes crucial.
Also, our ability to predict theoretically 
or measure in other contexts with some reliability the relative
values of the $f^d$'s becomes quite important.

\medskip
\noindent{\bf Estimates of the statistics required and the importance
of vetoing indirect decays.}
\smallskip

The ability to employ the $\Upsilon'/\Upsilon$ ratio
to analyze the nature of the QGP will depend crucially
on the statistics available. For $\sigma_{NN}T_{\rm PbPb}({\bf 0}) 
L_{\rm int}^{\rm PbPb}=4.973\times 10^3$~nb$^{-1}$ per month, 
as assumed in Table~\ref{tablerates},
and the $d\sigma(pp)/dp_T\times S(p_T)$ spectra
of Fig.~\ref{figuredsigdpt}, which range very roughly 
from 1 nb/GeV to 0.1 nb/GeV over
the $p_T$ range of interest, we find $\sim 5\times 10^3$ falling to
$\sim 5\times 10^2$ events per GeV
for the individual $\Upsilon$, $\Upsilon'$ and $\chi_b$
resonances. For branching ratios of $B_{\Upsilon}=0.025$
and $B_{\Upsilon'}=0.013$ and an efficiency/acceptance factor of 33\%
(which, as noted earlier, applies to the CMS detector), 
this leaves us with
$\sim 41.3$ ($\sim 21.5$) falling to $\sim 4.13$ ($\sim 2.15$) 
$\Upsilon$ ($\Upsilon'$) events per GeV per month
of running. For a 2 GeV bin size, 12 months of running
and negligible background (see below),
we obtain $\Upsilon$ ($\Upsilon'$) statistical accuracies 
of $\sim 3\%$ ($\sim 4.5\%$) at low $p_T$ rising to $\sim 10\%$ ($\sim 14\%$)
at $p_T\sim 25-30$ for the individual spectra, 
and of 5.4\% rising to 17\% for the $\Upsilon'/\Upsilon$ ratio.
Thus, if the prompt ratio can be extracted, somewhat more than
a year of running at $L_{\rm int}^{\rm PbPb}=2.59\nbi$ per month
would be sufficient to measure
the ratios with the required accuracy over the full
range of $p_T$ values of interest. If the $\Upsilon'/\Upsilon$ ratio
can only be measured including indirect decays, greater accuracy is
required at large $p_T$.  Roughly 36 months of
running at $L_{\rm int}^{\rm PbPb}=2.59\nbi$ per month would
be required to achieve $\lsim 10\%$ accuracy for the ratio at
$p_T\sim 30\gev$. Since this amount of running in the heavy ion
mode at the LHC is quite unlikely, we see how important the
vetoing of indirect decays is if we are to use the $\Upsilon'/\Upsilon$
ratio to analyze the QGP.  

\medskip
\noindent{\bf The importance of mass resolution.}
\smallskip

Also crucial to the program is the ability to resolve the $\Upsilon$,
$\Upsilon'$ and $\Upsilon''$ resonances from one another.
The CMS detector provides excellent mass resolution in the $\mupmum$ mode,
of order $40\mev$, adequate for clean separation
of the $\Upsilon$, $\Upsilon'$ and $\Upsilon''$
signals (see Fig.~12.31b in Ref.~\cite{CMS}).
As also shown in \cite{CMS}, the excellent resolution implies
a very modest level of background under the $\Upsilon$ peaks.
Preliminary estimates of the mass resolution for the ATLAS detector are
less encouraging, of order $200-300\mev$ \cite{ATLAS},
for which separation of the resonances would not be possible.

\medskip
\noindent{\bf Other onium ratios.}
\smallskip

In the above discussion, we have focused on the $\Upsilon'/\Upsilon$ ratio.
Other ratios would provide complementary information.  However, the 
$\Upsilon''/\Upsilon$ ratio is a factor of 4 smaller [see eq.~(\ref{fds})]
at the direct, \ie prompt, level and statistics would be a problem.
The direct/prompt $\chi_b(1P)/\Upsilon$ ratio is similar to the
$\Upsilon'/\Upsilon$ ratio, the major difference being that the 
$\chi_b(1P)/\Upsilon$ ratio never decreases below the $pp$ level at high
$p_T$ in the SU$(N)$ case since $p_{T, {\rm max}}^{\chi_b} \approx p_{T, {\rm
max}}^\Upsilon$ (see Figs.~\ref{figuresptsun} and \ref{figuresptgas}). 
Further, if the $\chi_b(1P)$ radiative decays can be detected
with reasonable efficiency,
the statistical accuracy with which the $\chi_b(1P)/\Upsilon$ ratio
can be measured would be comparable to that for the $\Upsilon'/\Upsilon$ ratio,
thus making it a most valuable
addition.  This adds to the importance of being able to trigger 
efficiently on the decay photons so as to isolate these two direct ratios.

\medskip
\noindent{\bf Extracting individual survival probabilities 
and using {\boldmath $Z$} production to control systematics.}
\smallskip

So far, we have considered only ratios of onium rates.  
However, it will be important to know that suppression
of the general type expected is taking place in
the individual spectra contributing to these ratios.
It would be even more valuable if a quantitative determination
of $S(p_T)$ for the different resonances
were possible through a direct comparison of the
differences between the $pp$ spectra and the Pb+Pb
spectra at $5.5\tev$, as illustrated in Fig.~\ref{figuredsigdpt}.
In principle, the ratio of the properly normalized Pb+Pb spectrum
to the $pp$ spectrum for a given resonance simply gives 
$S(p_T)$,\footnote{As argued in the previous section, non-perturbative
resummation effects upon the $p_T$ distributions will cancel
out in such a ratio.}
{\it provided} there are no non-QGP effects 
that can also alter the Pb+Pb spectrum. However, it is very likely
that there will be some nuclear effects from shadowing, comovers
and absorption. Effects from comovers and absorption
cannot be disentangled except as described earlier.
However, there is some possibility of obtaining
additional information regarding shadowing.

The strength of shadowing depends upon the $x$ values at which 
the parton distributions are probed.
Thus, the RHIC $pA$ measurements of onia production, while helpful, 
cannot directly determine the full effect of shadowing at $5.5\tev$
where we wish to determine the presence of the QGP and measure $S(p_T)$. 
An independent test of the shadowing effect using measurements
at $\rts=5.5\tev$ would be of great value for checking our understanding
of shadowing and assessing the accuracy with which it can be unfolded
from the data so as to reveal the QGP effects directly.

\begin{figure}[htb]
\let\normalsize=\captsize   
\begin{center}
\centerline{\psfig{file=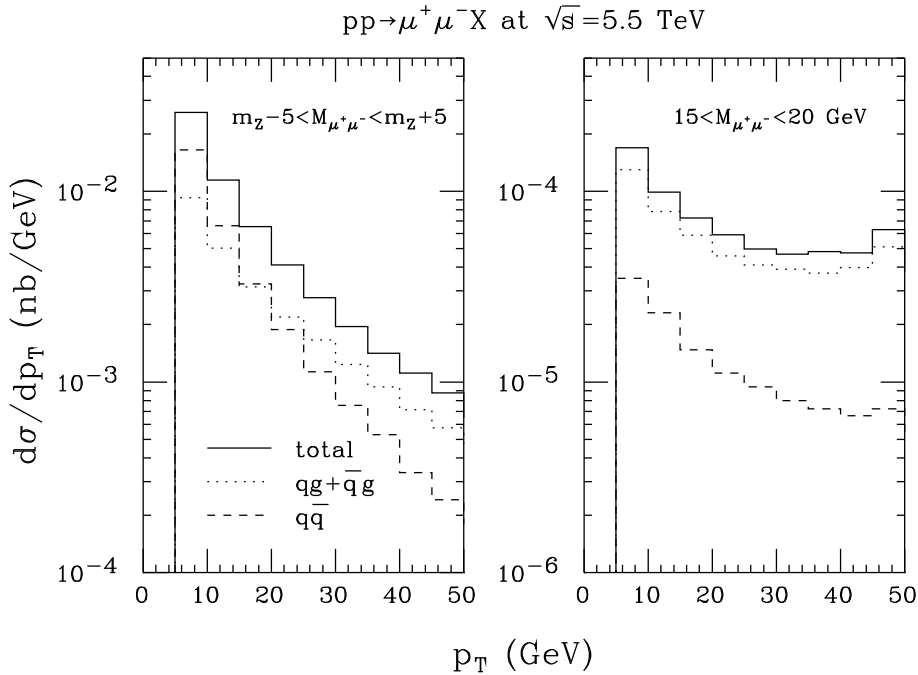,width=12.2cm}}
\begin{minipage}{12.5cm}       
\caption{In (a), we plot $d\sigma/dp_T$ for $pp\to \mupmum X$
at $\protect\rts=5.5\tev$ as a function of $p_T$ for $Z$ production
(defined by $\mz-5\gev\leq M_{\mupmum}\leq\mz+5\gev$). In (b),
the same cross section is plotted for $15\gev\leq M_{\mupmum}\leq 20\gev$.
The separate contributions from $q\anti q$ and $qg+\anti q g$ collisions
are indicated by dashed and dotted histograms, respectively.
}
\label{figurezpt}
\end{minipage}
\end{center}
\end{figure}

In the current experiments at the CERN SPS, 
$J/\psi$ production is compared to the dilepton continuum \cite{NA38,NA50}.  
The continuum is assumed to be produced via 
the Drell-Yan ($\gamma^\star,Z^\star$-exchange)
process and is, in fact, Drell-Yan-like in its behavior.  At RHIC energies and
higher, the continuum will be more difficult to understand because of the 
important contribution from semileptonic $c \overline c$ and $b \overline b$
decays.  Not only are there uncertainties in the total $c \overline c$ cross
section, but the heavy quark decays are also subject to nuclear effects.  The
relatively small Drell-Yan contribution is also subject to shadowing
effects in the mass range between the $J/\psi$ and the $\Upsilon$.
Another choice is needed.  One possibility
is $Z$ production. Because the $Z$ is produced
in point-like fashion, the difference
between the $p_T$-dependence of the $Z$ in $pp$ collisions
vs. Pb+Pb collisions will not be
influenced by the quark-gluon plasma\footnote{Once
again, this ratio will be very insensitive to low-$p_T$
perturbative and non-perturbative resummation modeling.}.

Figure~\ref{figurezpt} illustrates the cross section for $Z$ production
as a function of $p_T$ at $\rts=5.5\tev$ assuming no shadowing. 
There, we plot $d\sigma/dp_T$ for $pp\to ZX$ and the separate
contributions from $q\anti q$ and
$qg+\anti qg$ collisions.  (The distributions include 1-loop
corrections to the $Z+jet$ cross section\footnote{We thank
U. Baur for providing a program against which to check our
own calculations.},  but do
not include low-$p_T$ resummation effects.) To obtain the number
of events per month per GeV coming from central collisions\footnote{We
wish to restrict observations to central collisions in case
shadowing and nuclear effects (that might
possibly be present for the $Z$) for peripheral collisions are somewhat
different.}, we multiply by the factor
$\sigma_{NN}T_{\rm PbPb}({\bf 0})
L^{\rm PbPb}_{\rm int}=4.973\times 10^3$~nb$^{-1}$.
At $p_T=50$ GeV, the cross section
is of order $10^{-3}$ nb/GeV, implying about 2 events per GeV
for 40\% acceptance and detection efficiency. Thus,
for a 5 GeV bin we obtain about 10 events in this bin per month of running. 
After a year of running, this would yield a statistical accuracy of order 9\%.
At low $p_T$, event rates are a factor
of $\gsim 10$ larger, yielding correspondingly greater accuracy.  
The predicted effects of the QGP, see for example Figs.~\ref{figuresptsun}
and \ref{figuresptgas}, typically imply survival probabilities that
differ by much larger percentages compared to unity. In any case,
as estimated earlier, the errors in 
the measurements of the Upsilon resonance spectra will be larger.
Thus, production rates in the $p_T\lsim 50\gev$ domain are high 
enough that $Z\to \epem,\mupmum$ can provide a standard of comparison.
Of course, we will also wish to measure $Z$ production in $pp$
collisions at $\rts=5.5\tev$ so as to determine if shadowing and 
other nuclear effects have impacted the Pb+Pb spectra. Good statistical
accuracy for up to $p_T=50\gev$ requires integrated luminosity 
for $pp$ collisions of order
$L=0.01\fbi$, which should be easily achieved, even at this reduced energy,
by running for a few weeks.\footnote{If we assume
that low-luminosity running yields $pp$ integrated
luminosity of $10\fbi$ per year at $\rts=14\tev$
and that instantaneous luminosity falls as $E^2$, then the corresponding
integrated luminosity at $\rts=5.5\tev$ would be $\sim 1.5\fbi$
per year or $\sim 0.03\fbi$ per week.}

The two difficulties with using $Z$ production as a benchmark are: (i) $\mz$
is much larger than the masses of the Upsilon resonances and (ii)
the $q\anti q$ and $qg+\anti qg$ parton luminosities are probed
vs. the $gg$ luminosities responsible for $b\anti b$ production.
\begin{itemize}
\item
The large size of $\mz$ compared to the Upsilon family masses
reduces the value of $Z$ production as a benchmark for two reasons.
First, shadowing and related nuclear effects may well be dependent upon
$Q^2$. For example, in the limit of large $Q^2$, holding
the parton $x$ fixed, shadowing becomes stronger in some models
(see, \eg, Ref.~\cite{BrHe}
whereas in others it is predicted to disappear altogether
(see, \eg, Ref.~\cite{BKBP}). See also Ref.~\cite{NBH}.
Thus, it is very possible that the shadowing at
$Q^2=M_{\Upsilon}^2$ will differ substantially from that at $Q^2=\mz^2$.
Second, the $x$ values probed ($x \sim \mz/\rts \sim 0.016$ at $y=0$)
are much larger than in Upsilon production at the same energy.
(Detailed measurements as a function of the $Z$ rapidity would provide 
information about smaller $x$ values, 
but require more luminosity for good statistics.)
\item
The separate contributions of the $q\anti q$ vs. $qg+\anti qg$
densities to the $Z$ spectrum appear in Fig.~\ref{figurezpt}.
We observe that $q\anti q$ collisions are dominant for $p_T\lsim 15\gev$;
for higher $p_T$ values $qg+\anti qg$ collisions dominate.
Thus, to probe nuclear effects on the $g$ distribution
at $Q^2=\mz^2$, we must have knowledge of their effects
on the $q$ and $\anti q$ distributions for $x$
values in the vicinity of those being probed at this same $Q^2$.
At low to moderate $Q^2$ the nuclear effects on $q$ and $\anti q$
PDFs are rather well measured and do not appear to be 
strongly $Q^2$ dependent \cite{Arn}. However, there is currently
no direct measurement of $q$ and $\anti q$ shadowing at small
$x$ with $Q^2$ values as high as $\mz^2$. If nuclear beams
become available at HERA, then direct measurement of shadowing
at appropriate $x$ and $Q^2$ values would be possible.
\end{itemize}
Even though $Z$ production does not directly probe gluon 
shadowing at the $x$ and $Q^2$ values of interest for Upsilon
production,  $Z$ production would be of considerable value
in developing a reasonable theoretical understanding or model of shadowing.  
If we measure the $p_T$ dependence of
$Z$ production and find that it is consistent with one or more
theoretically reasonable models of shadowing, 
this will already be a very important benchmark
in that we can be confident there is not some totally
unexpected physics occurring in nucleon-nucleon collisions.
Of course, it is vital that $Z$ production be
measured both in Pb+Pb and in $pp$ collisions at $\rts=5.5$ TeV
in order for us to develop confidence that we understand the impact
of shadowing and other purely nuclear effects.

Given the above issues regarding $Z$ production,
it would be advantageous if we could measure lepton pair production
at pair masses nearer to those of the Upsilon states
so as to constrain shadowing and nuclear effects at parton $x$ and $Q^2$
values closer to those of direct relevance.  We have already
noted that we expect a large background from $c\anti c$ and $b\anti b$
production processes for lepton pair masses lower than
the Upsilon family masses. At these low masses, this background
will be very difficult to veto by requiring that the leptons
be isolated because of the high density of soft tracks
in the Pb+Pb collision environment.
In the mass region above about $15\gev$ the dilepton
rate from $c\anti c$ ($b\anti b$) pair production is predicted
to be smaller than (comparable to) that from $\gam^\star,Z^\star$-exchange
\cite{cleaner}. Further, in this higher mass range,
vetoing the $b\anti b$ component using isolation requirements
on the leptons might prove feasible at a level adequate
to extract the pure DY dilepton spectrum.
In Fig.~\ref{figurezpt}, we have plotted $d\sigma/dp_T$ for production
of muon pairs with $15\gev\leq M_{\mupmum}\leq 20\gev$
coming from $\gam^\star,Z^\star$-exchange. We note that
in this case the $qg+\anti qg$ collision component is always dominant,
as would be desirable for learning as much about gluon
shadowing as possible.  However, the cross section is nearly a factor
of 100 below that for production at the $Z$ resonance,
implying that statistics would be a factor
of 10 worse. Even a year of running will not provide
enough Pb+Pb luminosity to yield measurements that are sufficiently accurate
to constrain the shadowing and nuclear effects at the needed level ($\lsim
5-10\%$). Thus, the low rate and uncertainty regarding our ability
to veto the $b\anti b$ background imply that we should not count
on being able to use dilepton pairs below the $Z$ mass region to
improve our understanding of nuclear effects on the gluon PDFs.
Nonetheless, the possibility of doing so should not be ignored and
appropriate data, including event characteristics
that might allow vetoing, should be collected.

HERA measurements of the nuclear
$g$, $q$, and $\anti q$ densities as a function of $x$ and $Q^2$
would greatly assist in a reliable extraction of $S(p_T)$.  Measurements
overlapping the regions of interest for $\Upsilon$ and $Z$ production
at 5.5 TeV would be possible for an appropriate design of the facility.
Using the HERA measurements for $x\sim 0.02$ and $Q^2=\mz^2$,
it would be possible to
unfold the shadowing effects in $Z$ production and verify that
the underlying $p_T$ spectrum is that expected on the basis of
the Drell-Yan mechanism.  This would allow a high degree of confidence
that we could use the HERA measurements at the $x$ and $Q^2$
values appropriate to onium production to unfold the underlying
$p_T$ spectra of the onia as influenced only by absorption, comovers,
and the quark-gluon plasma, with the latter hopefully having a much
larger impact.

\section{Discussion and conclusions}

\indent\indent
We have found that expectations for $\Upsilon$ (and also $\psi$)
resonance production in Pb+Pb collisions at the LHC are very dependent upon
the nature and details of the quark-gluon plasma.  We have demonstrated
that, far from being an impediment to our extracting the QGP physics,
this dependence may very well allow us to determine much
about the fundamental nature of the QGP, including:
the energy density, the initial temperature, 
the plasma radius, and the temperature-dependent screening mass.
Although we have kept the discussion of quarkonium production and suppression
as general as possible, we chose the $\Upsilon$ family to examine in detail.
The $\Upsilon$ rate is high enough for statistically significant
measurements to be made, particularly at the LHC.  The $\Upsilon$ is also
less affected by interactions with hadronic matter, providing a less ambiguous
signal than the charmonium states.
Strikingly different expectations for the 
$p_T$-dependence of the $\Upsilon'/\Upsilon$ ratio,
the example upon which we focused, are found depending
upon whether or not the QGP is formed and on the QGP properties.
While our calculations of the survival probability
are rather schematic, {\it e.g.}\ assuming full equilibration,
they reflect the correct general trend for the scenarios discussed.

The much higher energy of the LHC could be crucial for QGP suppression of onium
production to be possible. This would be the case, for example,
in the minijet model of the initial conditions.
The high minijet density at the LHC implies 
that the $p_T$ spectra of the
$\Upsilon$ resonances in this model are likely to be highly sensitive to
details of QGP formation.  In contrast, at RHIC, the minijet density will be
a factor of 10 to 30 smaller, and the maximum initial temperature
predicted is $T_0\sim 0.5\gev$. Thus no suppression is expected.  Initial
conditions such as those predicted in parton gas models,
for which the QGP takes longer to equilibrate, 
predict still stronger suppression at the LHC and could also
lead to some suppression being observed at RHIC, depending
upon the temperature dependence of the screening mass in the $b\anti b$
potential. There are also models of the initial conditions that predict no
suppression even at the LHC.  We point out that even a very high $T_0$ does not
guarantee plasma suppression.  Models assuming a slow equilibration result in 
greater suppression because of the longer time the system spends in the
screening region with $T>T_D$.

Even if suppression is not present at RHIC, it is only at RHIC
that $pA$ measurements will be possible.  Despite 
the lower energy, these $pA$ measurements will greatly aid in 
constraining and checking our understanding 
of nuclear effects, such as
those associated with shadowing, comover interactions
and nuclear absorption. Thus, RHIC
measurements will, as a minimum, provide important benchmarks.

We have demonstrated that statistics should be adequate to
detect the differences in the $\Upsilon'/\Upsilon$
ratio that would discriminate between different QGP models
(assuming that the ratio can be measured at the prompt level).
Although this (and other) ratios have the advantage that
various systematic effects,
as well as the purely nuclear effects associated with shadowing,
will cancel, further vital information can be extracted if the $p_T$ spectra
for the individual resonances can be measured and the effects
of ordinary nuclear shadowing on these spectra unfolded.
We have pointed to a comparison with $Z$ production as a potentially
useful benchmark in this unfolding process. 
The ideal situation arises if, in addition to LHC $Z$ data, 
HERA nuclear data on shadowing is available for the $x$ and $Q^2$
values appropriate for both onium and $Z$ production at the LHC.
The observed $p_T$ spectrum for the $Z$ could be compared
to the prediction based on the Drell-Yan process
using the experimentally measured nuclear PDFs from HERA.
If there is agreement, then we would have a high
level of confidence that we can
unfold the shadowing effects on the onia $p_T$ spectra,
using the PDFs determined from HERA data at the appropriate $x$
and $Q^2$ values, so as to expose the impact of the QGP (which hopefully
will be dominant over absorption and comovers). Whether or
not the HERA data is available, $Z$ production will provide an important 
benchmark as to how much nuclear shadowing effects alone 
are likely to be impacting the measured $p_T$ spectra of the onia.

Certain detector features will play a key role
in carrying out the analysis as envisioned here. Of greatest
importance is the mass resolution in the $\mupmum$ channel.
The resolution for the CMS detector, as quoted in their technical
proposal, is such that the $\Upsilon$, $\Upsilon'$ and $\Upsilon''$
resonances can be cleanly separated from one another with very modest
background under the peaks.
Less certain is the extent to which the $\chi_b\to \gam\Upsilon$
process can be separated from direct $\Upsilon$ production by detecting
the associated $\gam$ for $p_T$'s in the range of interest. 
The statistics necessary to discriminate between different QGP models
is much greater if the direct `prompt' $\Upsilon'/\Upsilon$ ratio cannot
be isolated.  One also would lose the ability to observe the 
$\chi_b/\Upsilon$ ratio, which exhibits sensitivity to the QGP
model similar to $\Upsilon'/\Upsilon$, 
thus roughly doubling the significance with which
different QGP models can be differentiated.
It will be important for the LHC detector groups to
give further attention to both experimental issues.

\bigskip

\begin{center}
{\bf Acknowledgements} 
\end{center}

We would like to thank S. Brodsky, D. Denegri, W. Geist, W. Ko and P. Levai 
for discussions. 
We thank U. Baur for providing a $Z$ production program against
which to check our results. R.V. thanks the Institute for Nuclear Theory
at the University of Washington for its hospitality during
the completion of this work.

\vfill\eject


\begin{thebibliography}{99}
\bibitem{cleaner} These statements are based on the calculations of
Ref.~\cite{VGMR} for the LHC (using the acceptances of the ALICE experiment).
\end{thebibliography}
\end{document}